\documentclass[apj]{emulateapj}
\usepackage{graphicx,subfigure,amsmath, amsfonts, amssymb,aas_macros,times,epstopdf}
\usepackage[usenames,dvipsnames]{color}
\usepackage[export]{adjustbox}
\newcommand{\kms}{km~s$^{-1}$}
\newcommand{\cplus}{\hbox{[C~\sc{ii}]}}
\newcommand{\hi}{\hbox{H~\sc{i}}}
\newcommand{\oi}{\hbox{[O~{\sc i}]}}

\shorttitle{[C~{\sc ii}], [O~{\sc i}] and CO(1--0) in HCG 57}
\shortauthors{K. Alatalo et al.}

\begin{document}


\title{Strong Far-IR cooling lines, peculiar CO kinematics and Possible Star Formation Suppression in Hickson Compact Group 57}

\author{K. Alatalo,$^{1,2}$ P.~N. Appleton,$^{1,2}$ U. Lisenfeld,$^{3}$ T. Bitsakis ,$^{2,4}$ P. Guillard,$^{5}$ V. Charmandaris,$^{6,7,8}$ M. Cluver,$^{9}$ M.~A. Dopita,$^{10,11,12}$ E. Freeland,$^{13}$ T. Jarrett,$^{9}$ L.~J. Kewley,$^{10}$ P.~M. Ogle,$^{1}$ J. Rasmussen,$^{14,15}$ J.~A. Rich,$^{1,16}$ L. Verdes-Montenegro,$^{17}$ C.~K. Xu,$^{1,2}$ M. Yun$^{18}$}

\affil{$^{1}$Infrared Processing \& Analysis Center,  California Institute of Technology, Pasadena, CA 91125, USA\\
$^{2}$NASA Herschel Science Center, IPAC,  California Institute of Technology, Pasadena, CA 91125, USA\\
$^{3}$Departamento de F\'isica Te\'orica y del Cosmos, Universidad de Granada, Granada, Spain\\
$^{4}$Instituto de Astronom\'ia, Universidad Nacional Aut\'onoma de M\'exico, Aptdo. Postal 70-264, 04510, M\'exico, D.F., Mexico\\
$^{5}$Institut d'Astrophysique Spatiale, Universit\'e Paris-Sud XI, 91405 Orsay Cedex\\
$^{6}$Institute for Astronomy, Astrophysics, Space Applications \& Remote Sensing, National Observatory of Athens, GR-15236, Penteli, Greece\\
$^{7}$Department of Physics, University of Crete, GR-71003, Heraklion, Greece\\
$^{8}$Chercheur Associ\'e, Observatoire de Paris, F-75014, Paris, France\\
$^{9}$Astrophysics Cosmology and Gravity Centre, Dept of Astronomy, University of Cape Town, Private Bag X3, Rondebosch, 7701, Republic of South Africa\\
$^{10}$Research School of Astronomy and Astrophysics, Australian National University, Cotter Rd., Weston ACT 2611, Australia\\
$^{11}$Astronomy Department, King Abdulaziz University, P.O. Box 80203, Jeddah, Saudi Arabia\\
$^{12}$Institute for Astronomy, University of Hawaii, 2680 Woodlawn Drive, Honolulu, HI 96822, USA\\
$^{13}$The Oskar Klein Centre, Department of Astronomy, AlbaNova, Stockholm University, SE-106 91 Stockholm, Sweden\\
$^{14}$Dark Cosmology Centre, Niels Bohr Institute, University of Copenhagen, Juliane Maries Vej 30, DK-2100 Copenhagen, Denmark\\
$^{15}$Technical University of Denmark, Department of Physics, Building 309, DK-2800 Kgs. Lyngby, Denmark\\
$^{16}$Observatories of the Carnegie Institution of Washington, 813 Santa Barbara Street, Pasadena, CA 91101, USA \\
$^{17}$Departamento Astronom\'ia Extragal\'actica, Instituto Astrof\'isica Andaluc\'ia (CSIC), Glorieta de la Astronom\'ia s/n 18008 Granada, Spain\\
$^{18}$University of Massachusetts, Astronomy Department, Amherst, MA 01003, USA} 

\email{email:kalatalo@ipac.caltech.edu}

\begin{abstract}
We present \cplus\ and \oi\ observations from {\em Herschel} and CO(1--0) maps from the Combined Array for  Research in Millimeter Astronomy (CARMA) of the Hickson Compact Group HCG~57, focusing on the galaxies HCG~57a and HCG~57d. HCG~57a has been previously shown to contain enhanced quantities of warm molecular hydrogen consistent with shock and/or turbulent heating.  Our observations show that HCG~57d has strong \cplus\ emission compared to L$_{\rm FIR}$ and weak CO(1--0), while in HCG~57a, both the \cplus\ and CO(1--0) are strong. HCG~57a lies at the upper end of the normal distribution of \cplus/CO and \cplus/FIR ratios, and its far-IR cooling supports a low density warm diffuse gas that falls close to the boundary of acceptable PDR models.  However, the power radiated in the \cplus\ and warm H$_2$ emission have similar magnitudes, as seen in other shock-dominated systems and predicted by recent models. We suggest that shock-heating of the \cplus\ is a viable alternative to photoelectric heating in violently disturbed diffuse gas. The existence of shocks is also consistent with peculiar CO kinematics in the galaxy, indicating highly non-circular motions are present.  These kinematically disturbed CO regions also show evidence of suppressed star formation, falling a factor of 10--30 below normal galaxies on the Kennicutt-Schmidt relation.  We suggest that the peculiar properties of both galaxies are consistent with a highly dissipative off-center collisional encounter between HCG~57d and 57a, creating ring-like morphologies in both systems.  Highly dissipative gas-on-gas collisions may be more common in dense groups because of the likelihood of repeated multiple encounters.  The possibility of shock-induced SF suppression may explain why a subset of these HCG galaxies have been found previously to fall in the mid-infrared green valley.  
\end{abstract}





\section{Introduction}
Hickson Compact Groups (HCGs; \citealt{hickson82}) are small, relatively isolated systems of four or five galaxies in close proximity to one another.  These high density systems tend to have a high fraction of E/S0 galaxies compared with the field, show evidence of tidal interactions, and exhibit generally low group velocity dispersion \citep{hickson97}.  \citet{verdes-montenegro+01} studied the \hi\ properties in HCGs and found that they were often \hi\ deficient.  \citet{rasmussen+08} concluded that the \hi\ deficiency could not be explained solely by the heating of the gas.  \citet{borthakur+10} were able to show that a non-negligible fraction of the missing \hi\ might lie in an underlying diffuse component, but that diffuse component was unable to account for all missing \hi.  It was suggested that the level of \hi\ depletion could be seen as part of an evolutionary picture for the groups, evolving from more \hi\--rich systems composed of mainly spiral galaxies, towards groups with a higher fraction of elliptical and S0 galaxies \citep{konstantopoulos+10}.

\begin{figure*}[t!]
\subfigure{\includegraphics[height=2.7in,clip,trim=0cm 1.1cm 0cm 1.7cm]{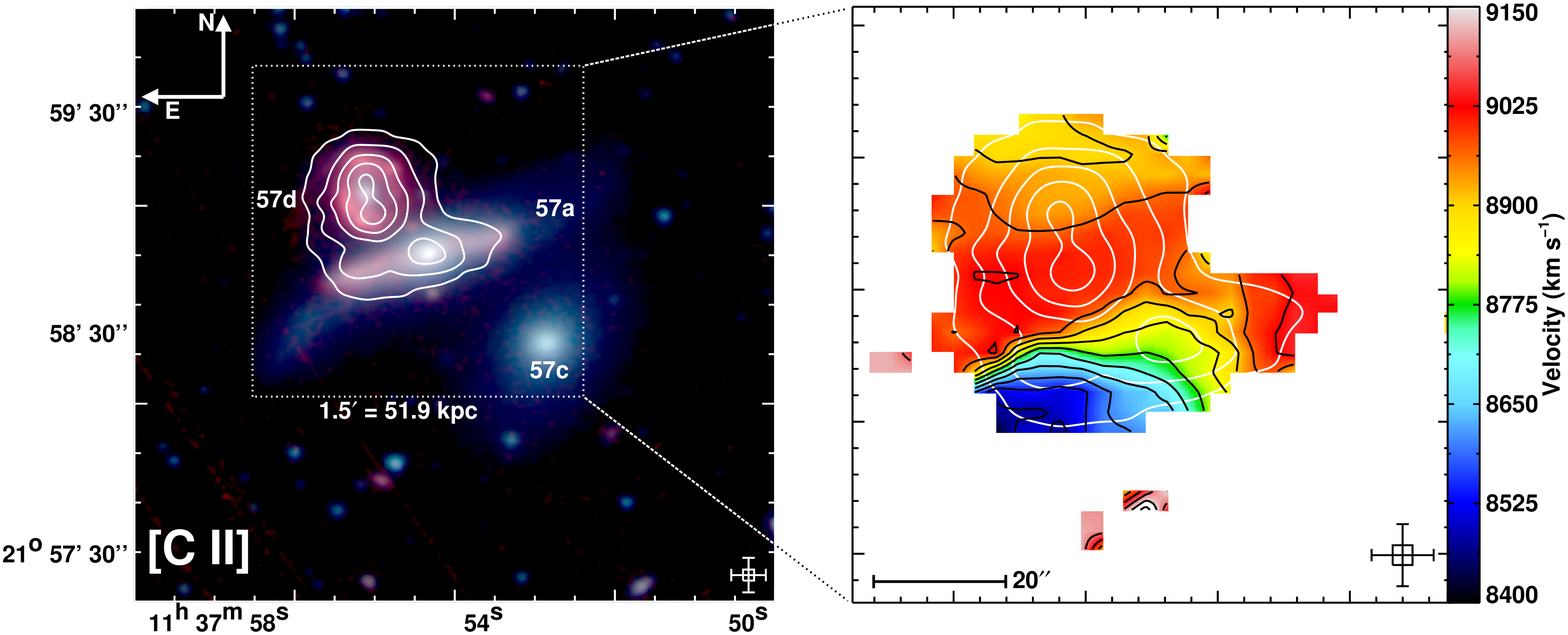}}
\subfigure{\includegraphics[height=2.7in,clip,trim=0cm 1.1cm 0cm 1.7cm]{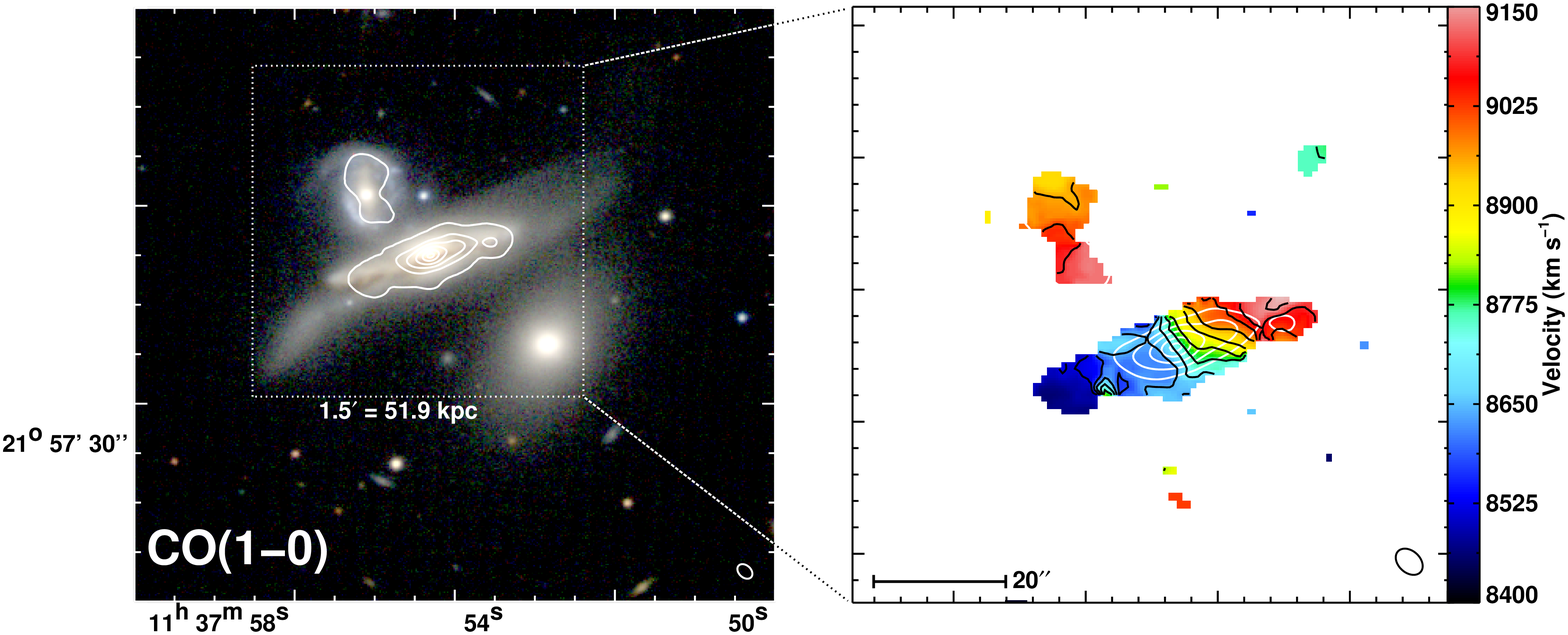}}
\caption{{\bf (Top):} The {\em Spitzer} 3.6\micron-4.5\micron-8.0\micron\ true color image (left, corresponding to blue, green and red, respectively) is overlaid with the integrated intensity contours (moment0; white) of the {\em Herschel} \cplus\ emission detected in HCG~57.  The dotted line box represents an area that is 1.5\arcmin\ on a side.  A zoomed-in map is shown on the right that includes the mean velocity (moment1) map of the \cplus\ data from {\em Herschel}, overlaid with isovelocity contours (black) as well as the moment0 contours (white). {\bf(Bottom):} The 3-color {\em g-r-i} SDSS image (left) is overlaid with the CO(1--0) moment0 map from CARMA (white contours).  A zoomed-in map is shown on the right which includes the moment1 map of the CO(1--0), overlaid with isovelocity contours as well as moment0 contours.  Major tick makes on both 3-color represent 1$'$, and 20$''$ on each moment1 map.  The CO(1--0) and \cplus\ velocity fields are consistent with one another, given the difference in spatial resolution between the two instruments, and some contamination of the \cplus\ velocity field of HCG~57a by 57d.  The largest mixing of kinematic components occurs in the souther-eastern part of the disk where the channel maps show the largest overlap of emission between the two galaxies. }
\label{fig:cii_co}
\end{figure*}

\citet{johnson+07} investigated the infrared (IR) colors of HCGs, using the  {\it Spitzer} IRAC diagnostic of \citet{lacy+04}. This work suggested that, unlike field galaxies,  HCG galaxies seem to display a ``gap'' in the number of galaxies of intermediate IRAC colors, lying between dusty IRAC-red galaxies, and dust-free stellar-dominated IRAC-blue galaxies--roughly corresponding to late-type and early-type galaxies respectively. This dearth in galaxies of intermediate color in HCGs, led to the idea that the environment of HCGs might somehow accelerate evolution from disks to early types \citep{tzanavaris+10,walker+10,walker+12,walker+13}.    \citet{bitsakis+10,bitsakis+11} studied the mid- and far-IR properties of HCGs, and found several trends in their specific star formation rates and dust properties consistent with the idea that enhanced galaxy interactions in HCGs drive the evolution of the global properties of the groups.  

Although the work of \citet{bitsakis+10,bitsakis+11} shows that in larger HCG galaxy samples, there is a small population of galaxies in the IRAC color ``gap" region, they tend to be galaxies which lie in the UV--optical green valley even when their colors are corrected for optical extinction. Furthermore, these galaxies may have unusual dust properties when compared with galaxies in the field \citep{bitsakis+14}. 

\begin{figure*}[t!]
\includegraphics[width=7.1in]{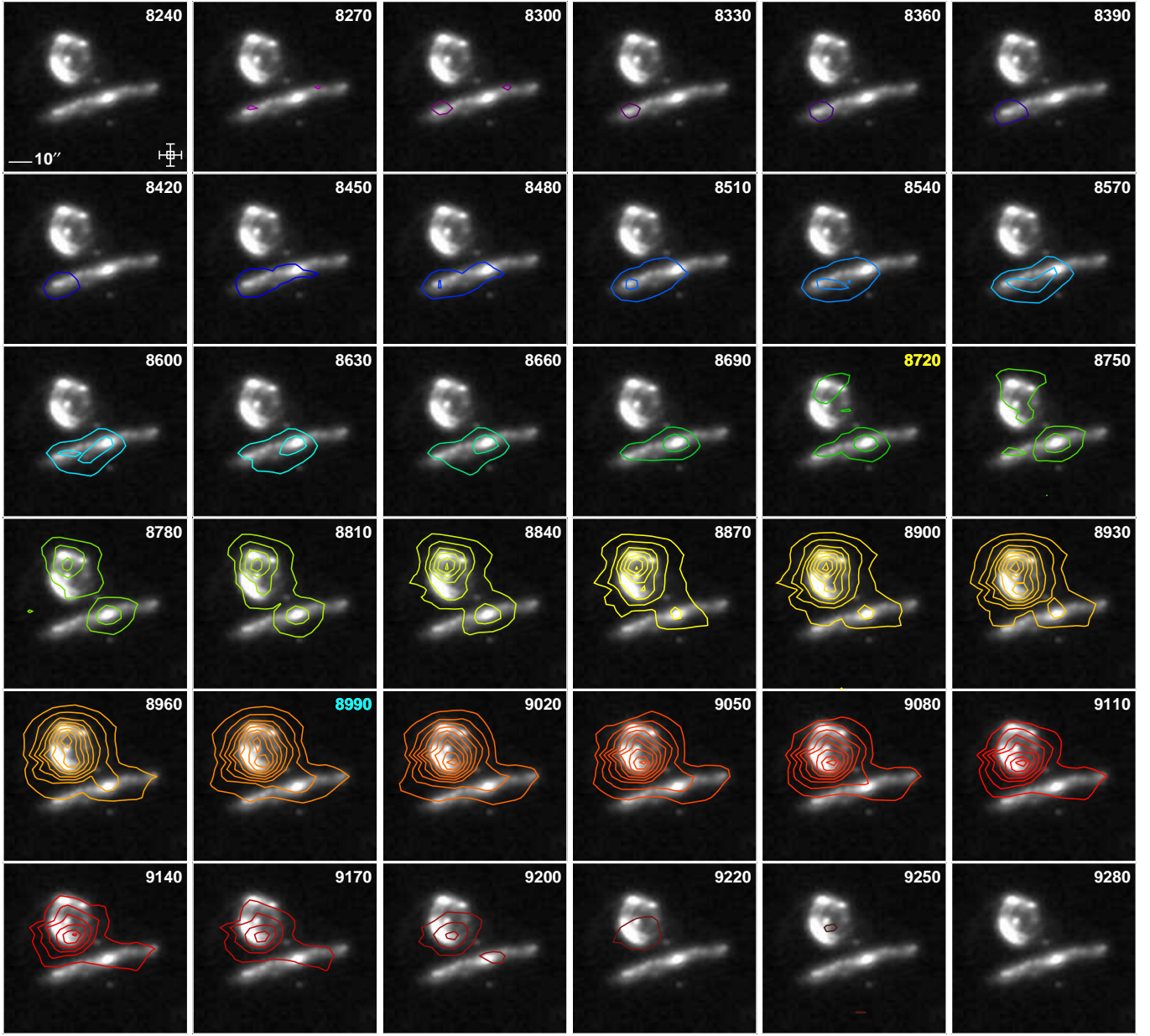}
\caption{The \cplus\ channel maps from {\em Herschel} listed in heliocentric, optically-defined velocities, with contour colors presented relative to the systemic velocity of HCG~57a overlaid on the {\em Spitzer} continuum-subtracted nonstellar 8.0\micron\ image. The \cplus contours colors represent the red- and blueshifted components seen.  Contours begin at $\pm3\sigma$ level and then are made in $3\sigma$ increments, with gray contours represent negative contours.  In the first panel we show a size scale (bottom left) and the 9\farcs4 Herschel PACS native spaxel is shown as a cross, overlaid with the chosen pixel size (3$''$) for the combined, projected \cplus\ maps (bottom right). \cplus\ is brightest in HCG~57d, but fainter emission is clearly present over a wide range of velocities in HCG~57A, especially near the nucleus.  Velocity labels with channels closest to the systemic velocities of HCG~57a (of 8727~km~s$^{-1}$) and of HCG~57d (of 8977~km~s$^{-1}$) are displayed in yellow and cyan, respectively.  The color of the contours consistently represents a velocity range between 8240--9330 \kms.}
\label{fig:cii_chans}
\end{figure*}

The mid-IR spectroscopic capability of {\it Spitzer} opened up the ability to study the pure rotational lines of molecular hydrogen in a large sample of galaxies. In one system, Stephan's Quintet (a.k.a. HCG~92) \citet{appleton+06} and \citet{cluver+10} discovered large quantities of intergalactic warm H$_2$ associated with a giant 40 kpc-scale shock in the system. Modeling of the observations suggests that strong turbulence was responsible for the formation of large quantities of warm H$_2$ from shocked H~{\sc i} gas \citep{guillard+09}.  The dissipation of turbulent energy is likely the main heating source of the molecular gas \citep{guillard+09,guillard+12a,appleton+13}. In a follow-up study, \citet{cluver+13} studied 78 HCG galaxies in 23 groups, and found that 20$\%$ showed unusually enhanced H$_2$ emission over and above that expected for heating by young stars alone ($I_{\rm H_2}$/$I_{\rm PAH(7.7\micron)}$ $\geq 0.04$).  Such galaxies, termed Molecular Hydrogen Emission Galaxies (MOHEGs) by \citet{ogle+07,ogle+10}, are also common in samples of nearby radio galaxies, where the H$_2$ is also likely powered by shocks \citep{guillard+12b}.  When the HCG MOHEGs  were placed on the IR color-color diagram, they were found to lie chiefly within the IRAC color ``gap'' leading \citet{cluver+13} to postulate that shocks and turbulent heating might be connected to their ``transitional'' IRAC colors.   

\begin{figure*}[t]
\subfigure{\includegraphics[height=2.1in]{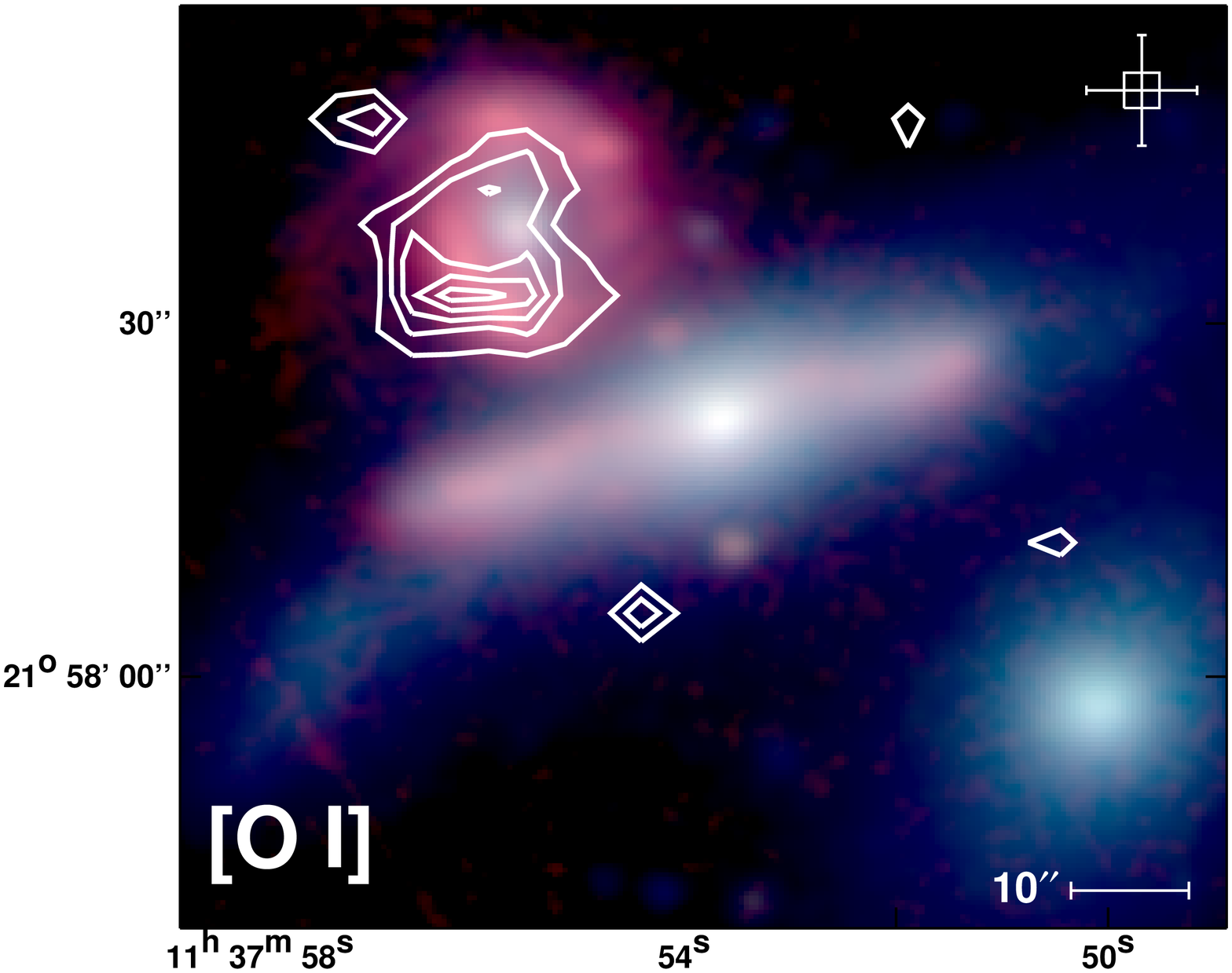}}
\subfigure{\includegraphics[height=2.1in]{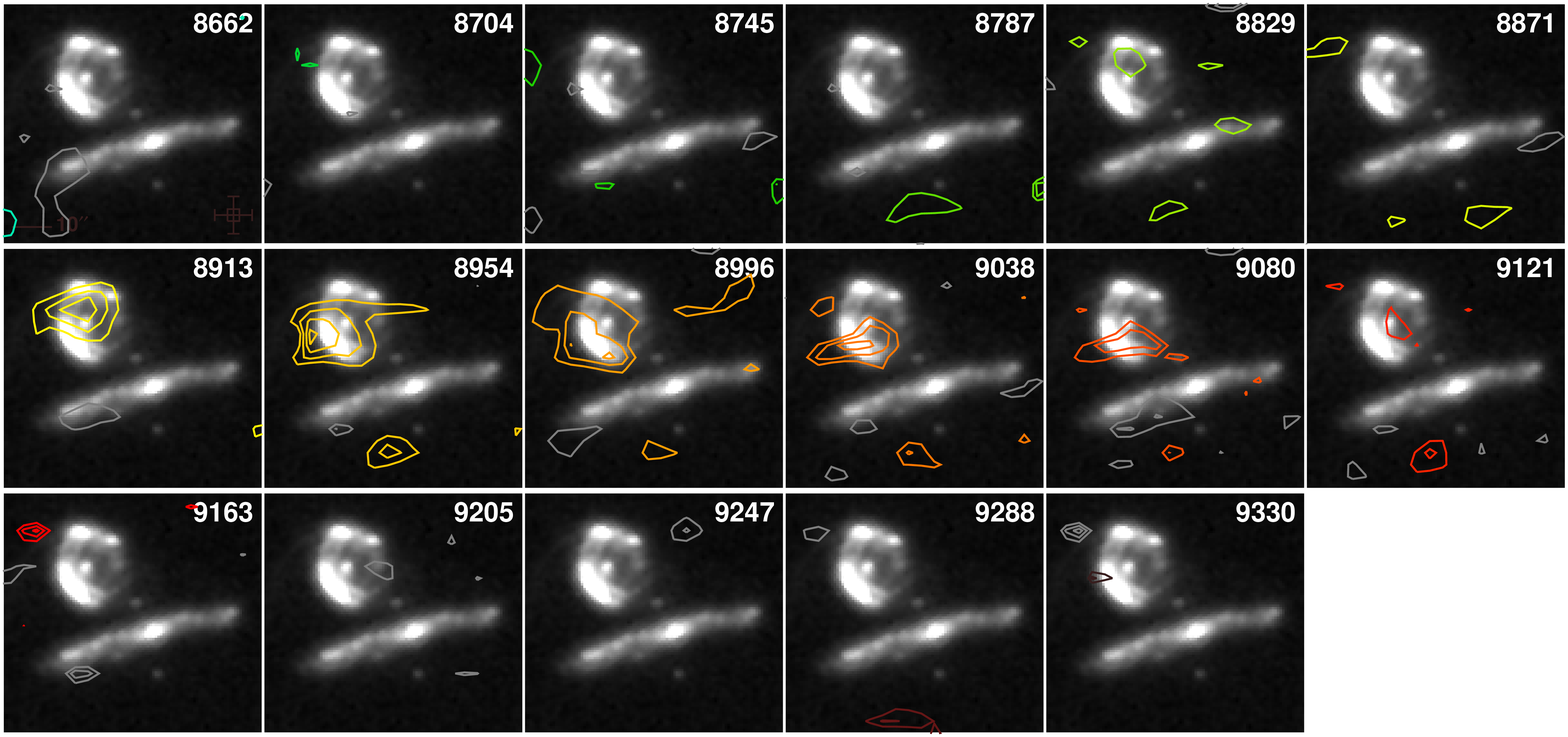}}
\caption{{\bf(Left):} The {\em Spitzer} 3.6\micron-4.5\micron-8.0\micron\ true color image (left, corresponding to blue, green and red, respectively) is overlaid with the integrated intensity contours (moment0; white) of the {\em Herschel} \oi\ emission detected in HCG~57, showing that \oi\ is primarily detected in HCG~57d.  {\bf(Right):} The {\em Spitzer} continuum-subtracted nonstellar 8.0\micron\ image is shown with the individual \oi\ channel maps from {\em Herschel} listed in heliocentric, optically-defined velocities, with contour colors presented relative to the systemic velocity of HCG~57a.  The colors of the contours represent the red- and blueshifted components seen.  Contours begin at $\pm2\sigma$ level and then are made in $1\sigma$ increments, with gray contours represent negative contours.  In the first panel and moment map we show a size scale (bottom left) and the 9\farcs4 Herschel PACS native spaxel is shown as a cross, overlaid with the chosen pixel size (3$''$) for the combined, projected \oi\ maps (bottom right). The \oi\ emission mainly follows the eastern side of the ring in HCG~57d, especially at higher systemic velocities.  The color of the contours consistently represents a velocity range between 8240--9330 \kms.}
\label{fig:oi_chans}
\end{figure*}

\begin{figure*}[t]
\includegraphics[width=7.1in,clip,trim=0cm 0cm 0cm 0cm]{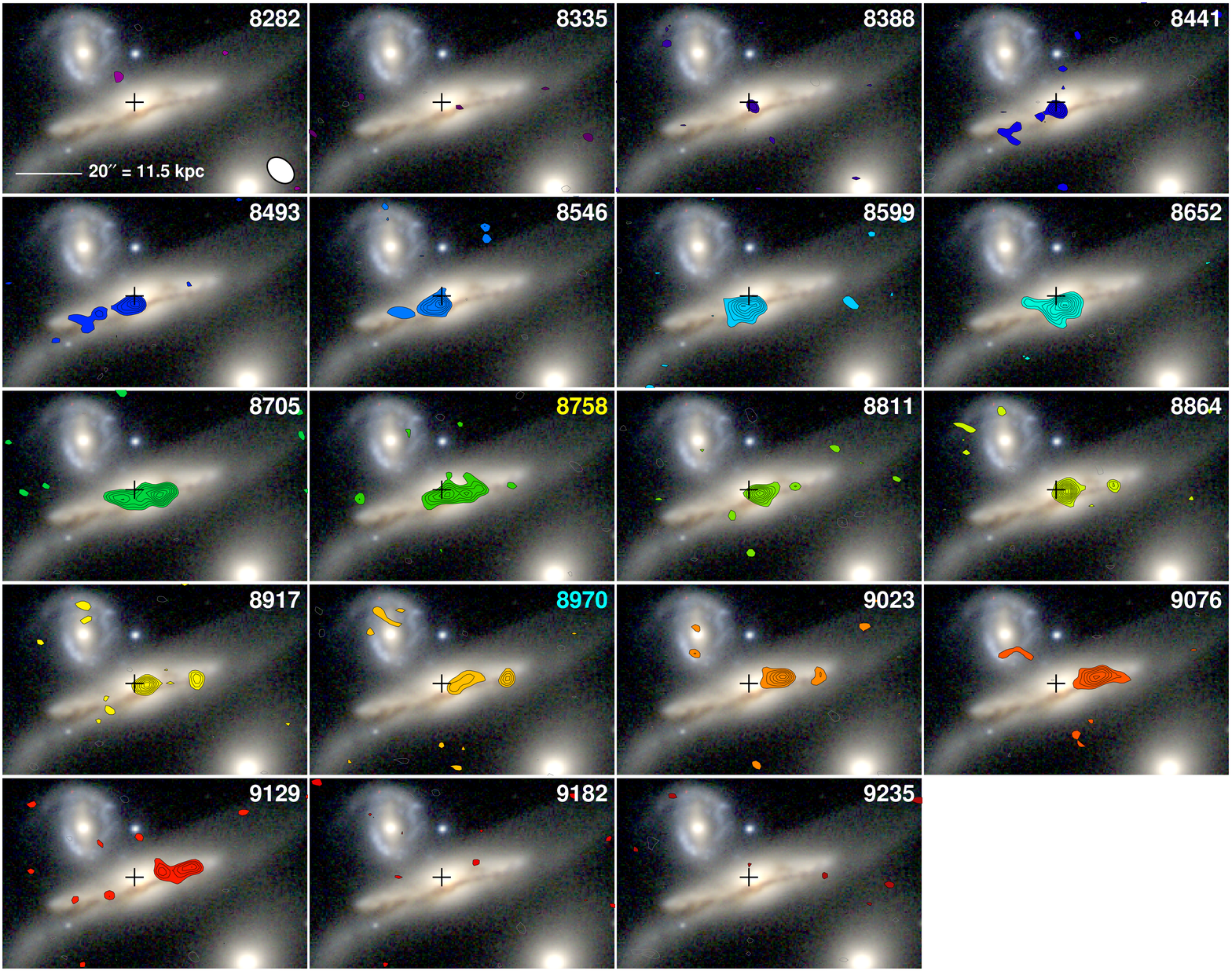}
\caption{The {\em g-r-i} SDSS image is shown with the individual CO(1--0) channel maps from CARMA, listed in heliocentric, optically-defined velocities with contour colors relative to the systemic velocity of HCG~57a.  The colors of the contours represent the red- and blueshifted components seen.  The black cross in each panel represents the nucleus of HCG~57a determined from the 2MASS imaging. Contours begin at $\pm2.5\sigma$ level and then are made in $1\sigma$ increments, with gray contours represent negative contours.  A size scale as well as the synthesized beam size of CARMA of $4.6\times3.3$ arcsecs are shown in the first panel.  
Channels closest to the systemic velocities of HCG~57a (of 8727~km~s$^{-1}$) and of HCG~57d (of 8977~km~s$^{-1}$) are listed in yellow and cyan, respectively.  The color of the contours consistently represents a velocity range between 8240--9330 \kms.}
\label{fig:co_chans}
\end{figure*}

Observations with {\em Herschel} \citep{herschel} have allowed us to further investigate these transition objects in other excitation tracers, such as \cplus\ and \oi.  Though \cplus\ preferentially traces photon-dominated regions (PDRs) in diffuse gas, and therefore correlates with star formation (the dominant energy source in most normal galaxies; \citealt{delooze+11}), \citet{appleton+13} have shown that in Stephan's Quintet, shocks are able to dominate the heating of the gas that gives rise to the \cplus\ emission.  Given the likelihood that galaxy interactions are more common at higher redshifts, it is important to gauge the importance of shocks and turbulence in enhancing \cplus\ emission from galaxies. 

It is also possible that the presence of shock-enhanced \cplus\ might be a signpost for {\em suppressed} SF.  \citet{ogle+10}, \citet{nesvadba+10} and \citet{guillard+14} show that in MOHEG radio galaxies, SF appears to be suppressed, possibly due to the turbulence induced by shocks as the radio jet traverses the molecular disk. \citet{guillard+12a} and \citet{konstantopoulos+14} conclude that SF is suppressed in the shock region of Stephan's Quintet.  On the other hand, simulations (as well as observations of ULIRGs; \citealt{sanders+mirabel96}) seem to show that SF can be enhanced during a galaxy collision \citep{saitoh+09,teyssier+10,bournaud+11}.  The current simulations do not reach the dynamical range needed to probe both the large-scale injection of mechanical energy and the dissipation scale where low-velocity shocks or vortices dissipate the energy. This is why in simulations the role that shocks play in regulating star formation is not yet completely understood. Even stronger suppression of SF has recently been found in the early-type galaxy NGC~1266 \citep{alatalo+14}, where suppression may be connected with an AGN outflow.   Therefore, getting a census of the molecular gas, \cplus\, and other tracers of SF (such as the 24\micron\ dust emission; \citealt{calzetti+07}) for a sample of likely shock-excited early-type HCG galaxies may help us understand the connection between \cplus\ and SF in turbulent galaxies, and provide insight into the potentially more turbulent era of high-z galaxies.

Here we present \cplus\ and \oi\ maps from {\em Herschel}, and CO(1--0) maps from the Combined Array for Research in Millimeter Astronomy (CARMA) for the inner part of the  HCG~57, which contains HCG~57a (= NGC~3753; a MOHEG) and its companions HCG~57d (= NGC~3754) and HCG~57c (= NGC~3750). Although this system is among the most \hi\ deficient of the groups studied by \citet{verdes-montenegro+01}, it contains copious quantities of warm molecular hydrogen. The rotationally excited H$_2$/PAH(7.7) ratio of HCG~57a is 0.17 \citep{cluver+13}, which is the third highest of the entire HCG MOHEG sample, placing it clearly outside the range of H$_2$ excitation that can be explained by photoelectric heating alone.  HCG~57a was also one of the few HCG galaxies observed by \citet{cluver+13} to show significant extended warm H$_2$ along its major axis, with M(H$_2$)$_{w}$ = 1.8$\times$ 10$^8$ M$_{\odot}$ of warm (T = 206 K) H$_2$ (over an area of 413~arcsecs$^2$, or 169~kpc$^2$). HCG~57c was not included as targets by the {\em Spitzer} Infrared Spectrograph (IRS) observations,, and although there was some coverage of HCG~57d, there was no clear signal of warm H$_2$, and thus little is known about their warm H$_2$ properties. Both HCG~57a and d were detected with the Institut de Radioastronomie Millim\'etrique (IRAM) 30m in CO(1-0) \citep{lisenfeld+14}.  

We intend to put HCG~57a and 57d forth as an example of the many ways that shocks and an interaction can influence the interstellar medium (ISM) and star formation (SF) within individual galaxies.  We adopt a distance to HCG~57 of 132~Mpc, based on the luminosity distance of the most massive component, HCG~57a, and a Hubble constant of H$_0$ = 67.3 km s$^{-1}$ Mpc$^{-2}$ and an updated $\Lambda$CDM cosmology with $\Omega = 0.315$, following the Planck parameters \citep{planck_cosmology}.  In \S\ref{obs}, we describe the observations and data reduction from {\em Herschel} and CARMA of HCG~57a and HCG~57d.  In \S\ref{results}, we compare the derived properties of each galaxy, including the spectra and fluxes of CO(1--0), \cplus, and \oi.  In \S\ref{discussion}, we discuss the properties of each galaxy individually, and compare those to normal galaxies, and build a picture of an encounter that might have created this system.  In \S\ref{summary}, we summarize our results.


\section{Observations}
\label{obs}

\begin{figure*}[t!]
\centering
\includegraphics[width=\textwidth,clip,trim=0cm 0cm 0cm 0cm]{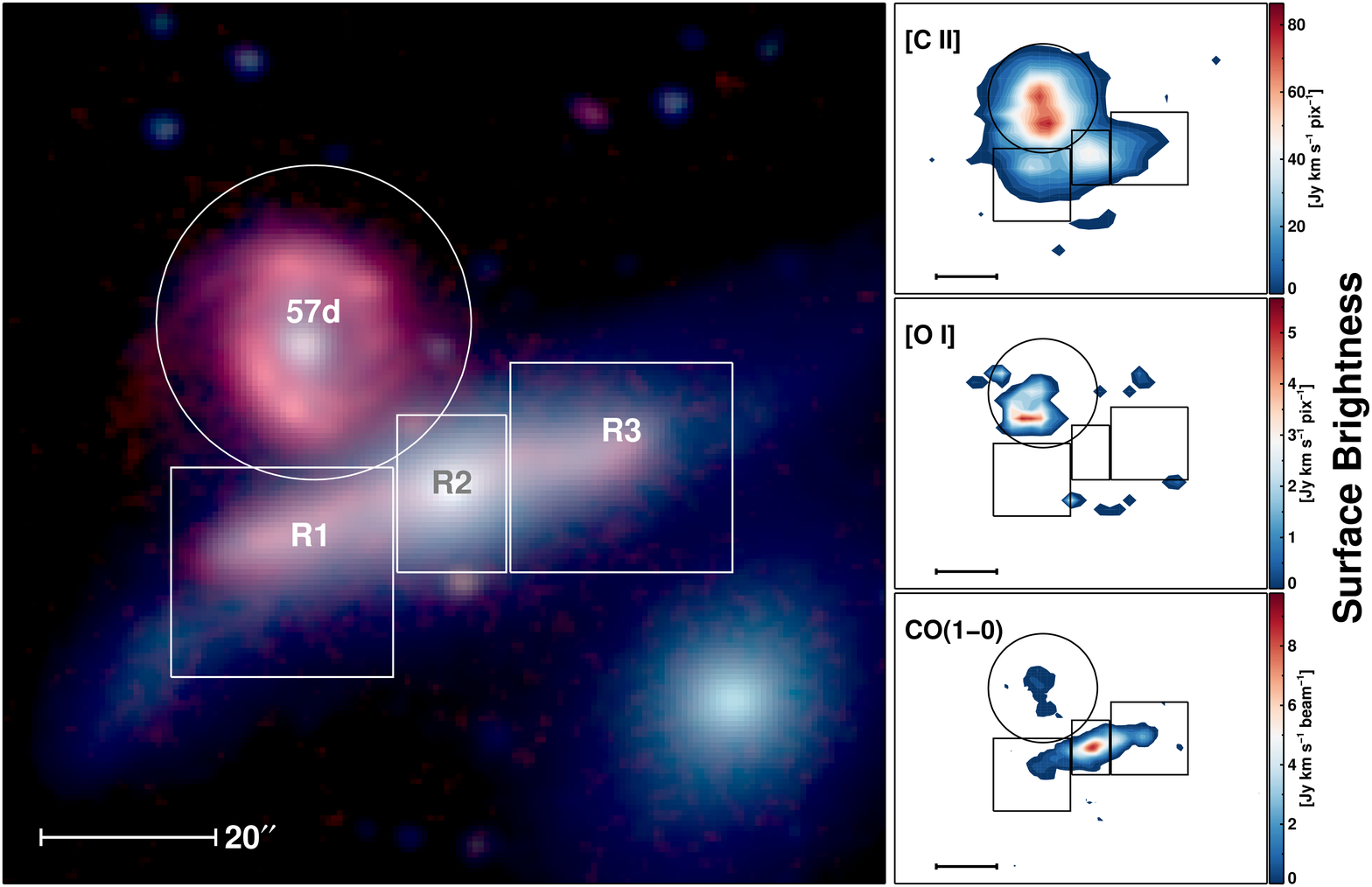}
\caption{The 3-color {\em Spitzer} IRAC 3.6, 4.5 8.0$\mu$m true color image (left) shows the demarcated regions on HCG~57a, as well as the circular region used to extract spectra for HCG~57d.  {\bf(Right)}  The smoothed integrated intensity maps in units of surface brightness of \cplus, [O~{\sc i}], and CO(1--0) are also shown with the regions overlaid.}
\label{fig:regions}
\end{figure*}

\subsection{Herschel}
{\em Herschel} observations with the Photodetector Array Camera \& Spectrometer (PACS; \citealt{pacs}) were made on 2012 June 29 and 2012 July 5 of the inner HCG~57 group covering HCG~57a/c/d in a $3\times3$ mapping mode with a map step size of 38 arcsecs. The observations were made as part of a {\em Herschel} open-time program (OT2$\_$pappleto$\_$2, PI: P. N. Appleton), and performed sparse mapping over an area $123\times123$ arcsecs$^2$.  The spectrometer targeted the redshifted \cplus$\lambda$157.74$\mu$m and \oi$\lambda$63.18$\mu$m lines, providing effectively 1408 s and 2464 s of integration time at each of the 9 pointings. Two ``range-mode''grating scans were obtained, covering a larger radial velocity range for the group (2600 km~s$^{-1}$ for \cplus\ and 1990 km~s$^{-1}$ at \oi) than a standard spectral-line observation.  The ``chop-nod'' mode was used with a (large) chopper-throw of 3 arcminutes, which ensured that the ``off'' position at each chopper position was clear of any galaxies. 

PACS data reduction was performed using the Herschel Interactive Processing Environment (HIPE) software package CIB13-3069 made available by the PACS team for our project.  Data processing, was performed on the level0 data (raw data) via the interactive pipeline, and included flagging (and ignoring) of bad pixels and saturated data, subtraction of chop ``on'' and ``off'' data, division of the relative spectral response function (RSRF), and the application of a flat-field. These data were converted from standard data frames to rebinned data cubes by binning these data in the wavelength domain using default parameters (oversample = 2, upsample = 4) which samples the spectra at the Nyquist rate in the two bands. Finally, data for the two nods were averaged, and final data cubes were created with 3$\times$3 arcsec$^2$ projected pixels on the sky using a new projection algorithm called "specInterpolate". This new algorithm which we obtained in advance of the full release of HIPE 13, will become the recommend projection algorithm of choice by the PACS team for this kind of PACS spectral mapping. Unlike previous projection methods (e. g. specProject) which simply divides up and averages flux from the PACS spectrometer spaxel measurement (9.4$\times$9.4 arcsec$^2$) onto the sky in a simplistic way, the new algorithm  uses a triangulation interpolation algorithm to more correctly distribute and average the various individual IFU pointings onto sky coordinates. We have extensively tested the algorithm for flux conservations with many different datasets, and confirm that the new and older (specProject) algorithms conserve flux  equally well. We assume that the absolute flux calibrations for the red \cplus\ and blue \oi\ final data cubes are uncertain by 12$\%$ and 11$\%$ respectively based calibration observations of standard calibrators.

Once channel maps of the {\em Herschel} \cplus\ and \oi\ were created, the moment maps were constructed in Interactive Data Language ({\tt IDL}).  The data cubes were Gaussian-smoothed spatially (with a FWHM equal to that of {\em Herschel} PSF at 160$\mu$m of 9.4 arcsecs), and masks were created by selecting all pixels above fixed flux thresholds, adjusted to recover as much flux as possible in the moment maps while minimizing the noise (generally about two to three times the rms noise in the smoothed channels). The moment maps were then created using the spatially-unsmoothed cubes within the masked regions, and integrating over the channel range where the mask indicated emission was detected. The {\em Herschel} PACS spectral resolution for \cplus\ was 230 \kms, and had a spectral step of 29.8~km~s$^{-1}$. The \oi\ maps were obtained with higher spectral resolution (90~km~s$^{-1}$) with a spectral step of 10.4~km~s$^{-1}$.

\subsection{CARMA}
Observations were carried out at the CARMA observatory between 2013 March 12 and 2013 March 17 in D-configuration (corresponding to baselines of 11-150 m, resolving scales between 3.6 and 48 arcseconds at \hbox{CO(1--0)}.  Observations included a long integration on a bright quasar (in this case, 3C273) in order to calibrate the passband response of the telescopes.  Observations were then taken in a repeating sequence, alternating between HCG~57a for 16 minutes, and a phase calibrator (the quasar 1224+213) for 2 minutes.  HCG~57a was observed for a total of 7.37 hours.  HCG~57d was within the primary beam of the HCG~57a field.  HCG~57c was also within the primary beam of the 6m CARMA antennas, but was undetected in CO(1--0).
\begin{table}[b!]
\caption{Flux fitting parameters for HCG~57}
\centering
\begin{tabular}{r l l l}
\hline \hline
{\bf HCG~57d} & {\bf CO(1--0)} & {\bf \cplus} & {\bf \oi} \\
\hline
$v_{\rm cen}$ (km~s$^{-1}$) = & $8976\pm13$ & $8977\pm4$ & $9000\pm28$ \\
FWHM (km~s$^{-1}$) = & $212\pm30$ & $309\pm9$ & $204\pm67$ \\
Peak Flux Density (Jy) = & $0.02\pm0.003$ & $12.0\pm0.31$ & $1.6\pm0.5$ \\
\hline
{\bf HCG~57a} & {\bf Region 1} & {\bf Region 2} & {\bf Region 3} \\
\hline
\multicolumn{4}{l}{\bf CO(1--0) Bandwidth (km~s$^{-1}$)} \\
8230--9342 & 8388--9023 & 8230--9183 & 8547--9278 \\
\hline
\multicolumn{4}{l}{\bf \cplus\ Bandwidth (km~s$^{-1}$)} \\
7910--9523 & 7910--9284 & 8120--9404 & 8537--9524 \\
\hline
\multicolumn{4}{l}{\bf \oi\ Bandwidth (km~s$^{-1}$)} \\
\multicolumn{4}{l}{\hskip 1cm 8306--9351 for all regions}\\
\hline \hline
\label{tab:params}
\end{tabular}
\end{table}

Raw CARMA data were reduced using the Multichannel Image Reconstruction Image Analysis and Display ({\tt MIRIAD}) package \citep{sault+95}, with the method of reduction and analysis identical to what is described in \S3.2 of \citet{alatalo+13}, and had a synthesized beam of $4.6\times3.3''$.  Data cubes were constructed with 30\kms\ channels of the central 90\arcsec, corresponding to the full-width at half-maximum (FWHM) of the primary beam of the 6m antennas.  Moment maps were then constructed using {\tt MIRIAD}, following the same method as is described in \citet{alatalo+13}.  


\section{Results and Analysis}
\label{results}

Figure \ref{fig:cii_co} shows the \cplus\ and CO(1--0) integrated distribution and mean kinematics, overlaid on both the infrared and the optical light.  Only the large edge-on disk galaxy (HCG~57a) and the small ring galaxy (HCG~57d; visible in {\em Spitzer} IRAC imaging) are detected by Herschel and CARMA: the elliptical galaxy HCG~57c is not detected.  The warp in the disk of HCG~57a suggests that it has been in a gravitational interaction, most likely with HCG~57d.  The 2-dimensional separation between the nuclei of HCG~57a and HCG~57d is 22\arcsec, or 13 kpc on the sky, and difference in the systemic velocities is 250~km~s$^{-1}$.   The CO(1--0) and \cplus\ velocity fields are consistent with one another, given the difference in spatial resolution between the two instruments, and some contamination of the \cplus\ velocity field of HCG~57a by HCG~57d.  The largest mixing of kinematic components occurs in the souther-easter part of the disk where the channel maps show the largest overlap of emission between the two galaxies. Bearing in mind this effect, the channel maps suggest that the CO and \cplus\ represent relatively well-mixed kinematic components.  A good example is the kinematics of HCG~57d, which in both \cplus\ and CO(1--0), show a gradient in the direction of HCG~57a, supporting the idea that this galaxy is interacting with the larger galaxy.  It is also possible that the \cplus\ channel maps in Figure \ref{fig:cii_chans} show a \cplus\ bridge between the galaxies, but the resolution and spaxel size of {\em Herschel} PACS are such that we cannot rule out a simple superposition of gas from the two galaxies in that region. 

\begin{figure*}[t!]
\subfigure{\includegraphics[width=\textwidth]{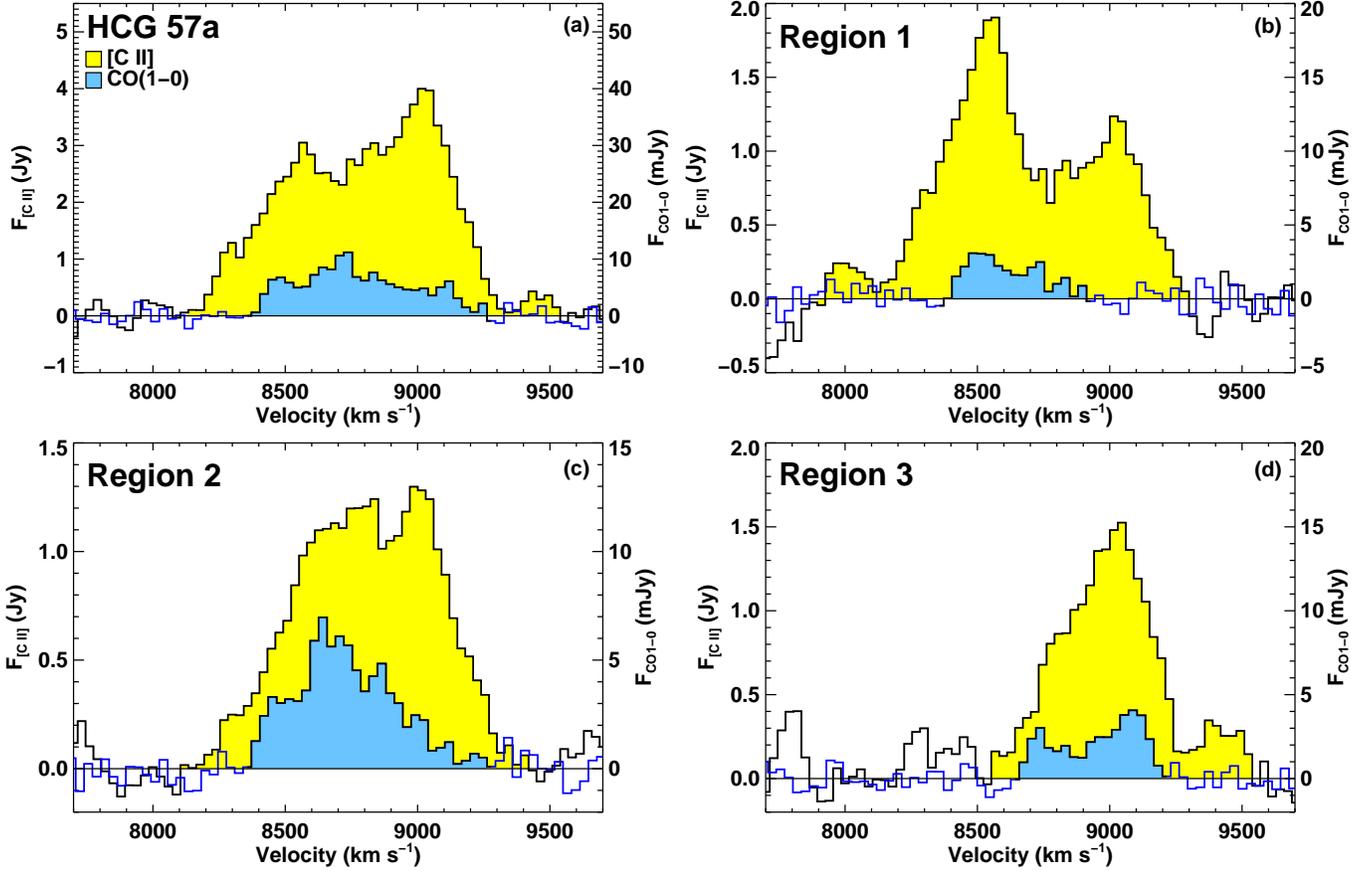}}
\caption{{\em Herschel} \cplus\ emission (yellow) compared to the CO(1--0) emission from CARMA (blue) in HCG~57a as well as the three sub-regions.  In most cases, the \cplus\ and CO(1--0) appear to trace gas of a similar velocity, except in the case of Region 1, in which the \cplus\ flux arising at $v \approx 9000$~km~s$^{-1}$ is likely contamination from HCG~57d. We estimate that less than 15\% of the \cplus\ emission from Region~2 is contamination from HCG~57d (see text).}
\label{fig:57a_specs}
\end{figure*}


Emission from the \oi$\lambda$63$\mu$m line was faintly detected from HCG~57d (and not at all from HCG~57a).  Unlike the emission from \cplus, Figure \ref{fig:oi_chans} shows that the \oi\ is more concentrated along the eastern edge of the galaxy in the region where the ring is especially prominent in near-IR observations. \oi\ emission is not detected in HGC~57a.


The integrated distributions of the \cplus\ and CO(1--0)  show large differences between the two galaxies.  HCG~57d, the dominant \cplus\ and \oi\ emitter in the system, is quite weak in CO(1--0), whereas the opposite is true for HCG~57a.   To further explore this effect we present the \cplus\ and CO(1--0) channel maps in Figures \ref{fig:cii_chans} and \ref{fig:co_chans}.  The \cplus\ channel maps (Fig. \ref{fig:cii_chans}) were overlaid on the {\em Spitzer} 8\micron\ non-stellar emission\footnote{This image was obtained by convolving the 3.6\micron\ image to the resolution of the 8$\mu$m observations, and then subtracting it from the 8.0\micron\ image with scaling factor of 0.232 \citep{helou+04}. This map is therefore dominated by PAH emission and warm dust (if present)}.  The channel maps confirm that \cplus\ is found throughout the disk of HCG~57a, but is brightest in HCG~57d.  CO(1--0) channels from CARMA are over-plotted on the 3-color SDSS image of HCG~57 (Fig. \ref{fig:co_chans}).  It is clear that there is much stronger CO emission in HCG~57a than HCG~57d, although weak (S/N per channel $\approx3$) emission is detected in HCG~57d, with a strong N/S velocity gradient--as discussed earlier.

\citet{lisenfeld+14} observed HCG~57a and 57d with the IRAM 30m.  Our CARMA observations recover 87\% of the IRAM 30m flux for HCG~57a and 51\% for 57d.  The 87\% flux recovery in HCG~57a is well within calibration errors for millimeter observations.  Table \ref{tab:params} lists the velocity bandwidth and fitting parameters used for each system, and Table \ref{tab:fluxes} shows the derived fluxes.  In HCG~57d, the under-recovery by the interferometer of flux could possibly be due to a large underlying mass of diffuse CO emission, however the overall faintness of CO in HCG~57d means that we are signal-to-noise limited on the scale of the synthesized beam of the interferometer for this galaxy.  Therefore, for the purposes of {\it global} properties of HCG~57d, we assume the 30m-measured CO fluxes, luminosities and masses.  When deriving properties such as star formation surface density (\S\ref{supp_SF}), because the area of the emitting source is important, we use the CARMA-derived areas and surface densities for HCG~57d.

In order to create data cubes suitable for spectral comparisons, we convolved the CO(1--0) map to the resolution of the \cplus\  data (9.4\arcsec).  The smoothed data cube was then used to make moment maps using {\tt MIRIAD} as described in \citet{alatalo+13}.  The resultant CO(1--0) maps were then registered to the same coordinate system as the \cplus\ map. \oi\ emission was only detected in HCG~57d.  The \oi\ data cube was registered to the coordinate system of the \cplus\ map, but was not convolved since the only emission detected (associated with HCG~57d) is quite compact, and well contained within the large extraction region centered on that region.   Spectra (described in the next subsections) were then extracted in four chosen regions: one centered on HCG~57d, and 3 regions covering HCG~57a.  All \cplus\ and \oi\ pixels from the registered cubes lying inside of the extraction regions of Fig. \ref{fig:regions} were summed to create the spectra.  An additional step was taken for the \cplus\ spectrum in Region 1 of HCG~57a, which likely included a contribution from the background HCG~57d.  Figure \ref{fig:57a_specs}b indicates that there is a large redshifted \cplus\ peak that is not also traced by CO(1--0) emission, which is the component suspected of being part of the contamination.

A two Gaussian fit was run on the Region~1 \cplus\ spectrum, and gas that appeared centered on $v_{\rm sys}$ of HCG~57d was measured to be $\approx 40$\% of the total \cplus\ signal from Region~1. The boundaries of Region~2 were chosen specifically to minimize \cplus\ contamination of HCG~57a from HCG~57d by careful inspection of the \cplus\ data cube. By extracting spectra from between the two galaxies, and estimating the extent of the emission "skirt" from HCG~57d in the relevant velocity channels towards HCG~57a,  we  believe the contamination of Region~2 is between 10 and 15$\%$ over its broad \cplus\ velocity profile. This relatively low total contamination is aided by the fact that the centroid of the main \cplus\ emission from HCG~57d shifts to the south-east at the velocities around 9000 km s$^{-1}$ and higher, largely avoiding Region~2. Furthermore, emission isolated to the central regions of HCG~57a is observable in a wide range of channels, even over the range where the two galaxies have similar velocities, again suggesting that the contamination is not large.  Strong contamination would pull the observed emission centroid away from the center of HCG~57a, which is not observed except perhaps in the channels between centered on 8960 and 8990 km s$^{-1}$.  Region~3 does not appear to suffer significant contamination.  The total \cplus\ luminosities for the galaxies and selected regions throughout this paper take these estimated contamination factors into account.  

\begin{table*}[t!]
\caption{HCG~57 \cplus, CO(1--0) and \oi\ Luminosities$^\ddagger$}
\centering
\begin{tabular*}{7.1in}{l| c c c| c c| c c c c}
\hline \hline
& L$_{\cplus}$ &L$_{\rm CO(1-0)}$&L$_{\oi}$ & L$_{\rm TIR}$$^\ast$ & L$_{\rm FIR}$$^\ast$ & \cplus/L$_{\rm FIR}$ & CO(1--0)/L$_{\rm FIR}$ & \cplus/\oi & \cplus/CO(1--0) \\
& $10^7\times$(L$_\odot$) &$10^4\times$(L$_\odot$)&$10^6\times$(L$_\odot$)&$10^9\times$(L$_\odot$)&$10^9\times$(L$_\odot$) & $\times10^{-3}$ &$\times10^{-6}$ & & \\
\hline \hline
HCG~57d & 12.5$\pm$0.51 & 
2.05$\pm$0.22$^\dagger$ & 
27.4$\pm$12. & 
21.1$\pm$5.6 & 
8.4$\pm$2.3 & 
15.$\pm$4.1&
2.4$\pm$0.7&
4.6$\pm$2.&
6100$\pm$690\\
\hline
HCG~57a & 7.99$\pm$0.18$^\diamondsuit$ & 
9.95$\pm$0.49 & 
$<$ 13.2 & 
23.3$\pm$4.2 & 
6.24$\pm$1.2&
13.$\pm$2.&
16.$\pm$3.1&
$>$ 6.1&
804.$\pm$43.\\
Region 1 & 2.47$\pm$0.14$^\diamondsuit$ & 
1.83$\pm$0.24 & 
$<$ 8.6 & 
8.65$\pm$2.8 & 
2.5$\pm$0.87&
9.8$\pm$3.5&
7.3$\pm$3.&
$>$ 2.9&
1350$\pm$190\\
Region 2 & 2.48$\pm$0.09 & 
5.49$\pm$0.22 & 
$<$ 5.0 & 
7.9$\pm$2.0 & 
1.7$\pm$0.5&
14.$\pm$4.4&
31.6$\pm$9.9&
$>$ 5.0&
452.$\pm$25.\\
Region 3 & 2.17$\pm$0.09 & 
2.45$\pm$0.17 & 
$<$ 7.3 & 
6.7$\pm$1.9 & 
2.0$\pm$0.6&
11.$\pm$3.2&
12.$\pm$3.6&
$>$ 3.0&
887.$\pm$73.\\

\hline \hline
\label{tab:fluxes}
\vspace{-3mm}
\end{tabular*}
\raggedright $^\ddagger$assuming a luminosity distance D = 132~Mpc\\
$^\ast$L$_{\rm FIR}$ is 42.5--122.5$\mu$m \citep{helou+88} and L$_{\rm TIR}$ is from 8--1000$\mu$m, values from \citet{bitsakis+14}.\\
$^\dagger$from \citet{lisenfeld+14}, $L_{\rm CO} = 1.07\pm0.22\times10^4$ L$_\odot$ recovered from CARMA\\
$^\diamondsuit$includes a correction for \cplus\ contamination from HCG~57d of 9\% for the total HCG~57a system and 32\% for Region 1, 15\% for Region~2, and no contamination for Region~3.
\end{table*}

Due to the low S/N nature of the CO(1--0) cube, an extra step was added to CO spectral extractions.  The pixels with non-zero emission in the smoothed CO(1--0) moment0 map inside the region were the only ones considered in making the spectrum.  This avoided adding large, noise-dominated areas to the integral, which would dilute the CO(1--0) signal.  The right panels in Figure \ref{fig:regions} identify the extractions that were used in making the spectra of \cplus\ and CO(1--0) for each region.

\subsection{HCG~57d}
The CO(1--0) distribution and kinematics within HCG~57d appear to represent emission originating from the northern and southern parts of a face-on ring (which is most obvious in the non-stellar 8$\mu$m emission seen in the top left panel of Figure \ref{fig:cii_chans})  The CO(1--0) kinematics (seen in Figures \ref{fig:cii_co}b \& \ref{fig:co_chans}) in HCG~57d also appear fairly regular, and in general follow the kinematics of the brighter (but also less well-resolved) \cplus\ emission.  As discussed previously, the \oi\ emission seems to follow the eastern and south-eastern segment of the star forming ring as seen in \ref{fig:oi_chans}.

\begin{figure}[t!]
\includegraphics[width=0.49\textwidth]{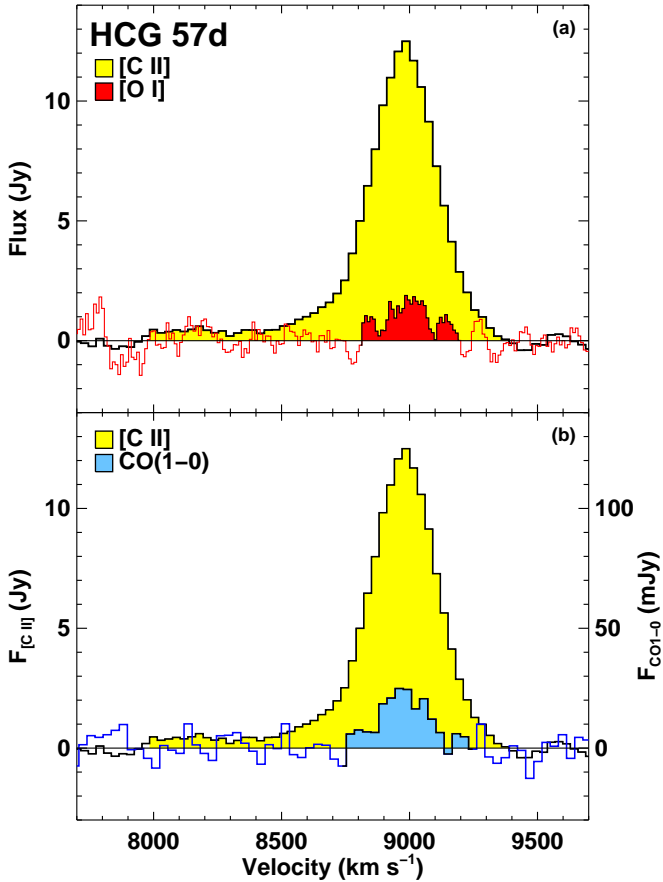}
\caption{{\bf (Top)} {\em Herschel} \cplus\ emission in HCG~57d (yellow) compared to the [O~{\sc i}] emission (red).  Both \cplus\ and [O~{\sc i}] are detected in HCG~57d. The ratio (Table 2) of the integrated \cplus\ to \oi\ line fluxes is on the high end of the distribution for normal galaxies (see text).  {\bf(Bottom)}  The \cplus\ spectrum of HCG~57d is compared to the CO(1-0). Again the integrated flux ratio of CO to \cplus\ is unusually low compared with normal galaxies.}
\label{fig:57d_specs}
\end{figure}

The integrated CO(1--0), \cplus, and \oi\ spectra for HCG~57d are shown in Figure \ref{fig:57d_specs}. The spectra are quite similar, suggesting that the atomic gas and  molecular gas are well mixed, although the \oi\ have a slightly higher systemic velocity than the \cplus\ and CO, perhaps because it has a slightly different distribution from the \cplus.  In order to derive the integrated line fluxes, each line was fit by a single gaussian profile using IDL code {\tt gfitflex}\footnote{http://astro.berkeley.edu/$\sim$heiles/handouts/handouts\_idl.html}.   Table \ref{tab:params} presents the fitting parameters for the gaussian fits for CO(1--0), \cplus\ and \oi\ for HCG~57d.  Table \ref{tab:fluxes} presents the results, and their total errors\footnote{A sum of the fitting errors and the root mean square noise in the individual pixels}.   The \cplus/CO(1--0) ratio is $\approx 6100$, larger than is seen in most galaxies ($\approx 1000$; \citealt{malhotra+01}), though is consistent with the extreme starburst from \citet{stacey+91} and high-$z$ galaxies from \citet{stacey+10} (with an \cplus/CO(1--0) average of $\sim6000$).  The extreme environments of the starbursts and submillimeter galaxies support extremely large ultraviolet fields (log G$_0>3$), which is inconsistent with the observations of HCG~57d, discussed in more detail in \S\ref{disc:pdr}.

\subsection{HCG~57a}
\label{res:HCG57a}

\begin{figure*}[t]
\subfigure{\includegraphics[width=\textwidth]{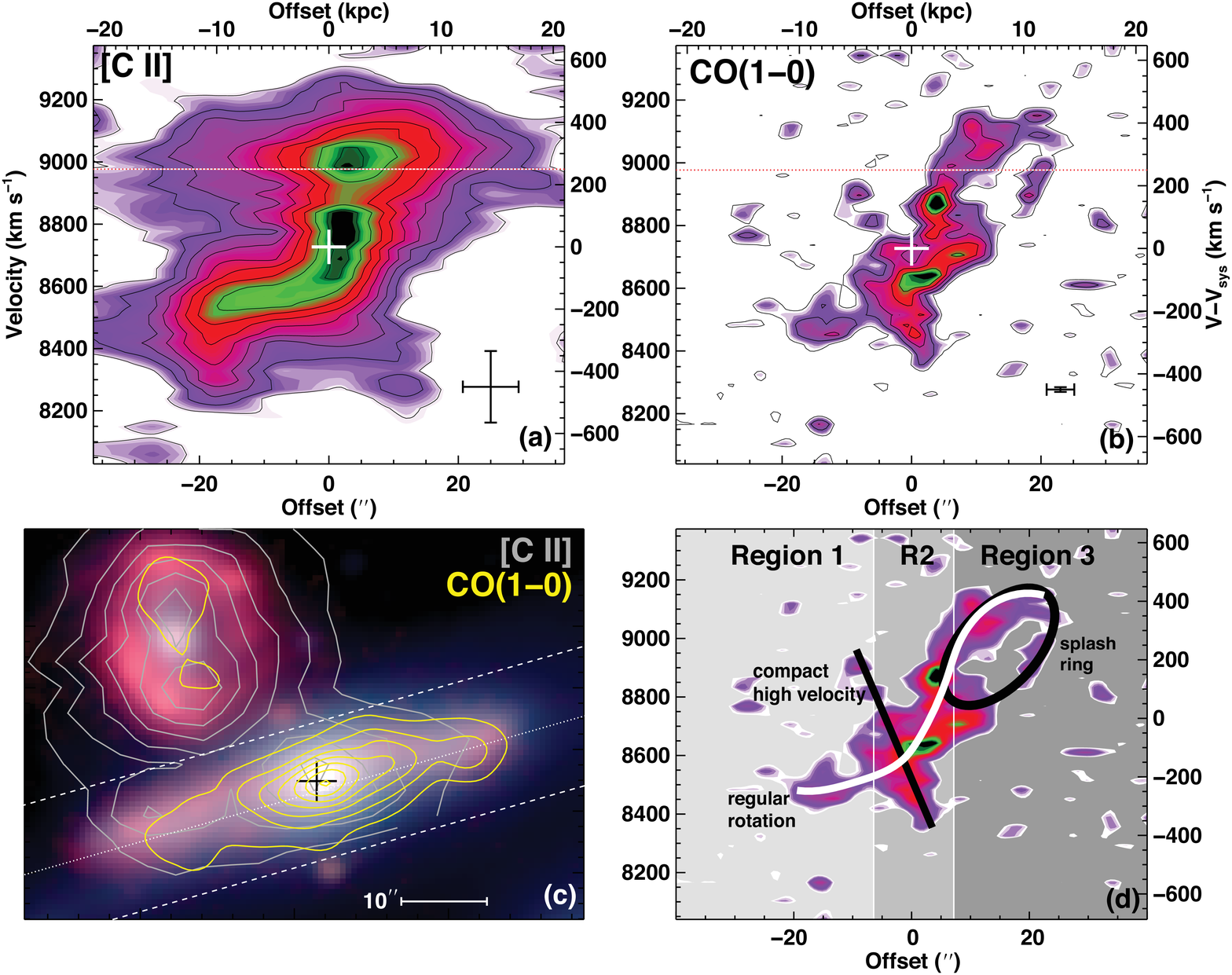}}
\caption{{\bf(Top left):} Position-velocity diagram (PVD) in \cplus\ of HCG~57a. Although it is clear that smoothing both of the pixels as well as the velocities is an issue, the PVD shows multiple components.  There seems to be a \cplus\ emission component originating at the systemic velocity of HCG~57d, but is only offset by $\approx$5$''$ from the HCG~57a nucleus.  {\bf(Top right):} The CO(1--0) PVD, which has been derived from the same spatial box as the \cplus, though was less extended.  The red dotted horizontal line appears in both PVDs and represents the systemic velocity of HCG~57d.  The white point represents the velocity and RA center of HCG~57a, as determined from a 2MASS $K_s$-band image.  The black points show the uncertainties in each axis.  The position of the minimum and maximum velocity offset for the contours, which for CO(1--0) are -390 and 465 km s$^{-1}$, and for \cplus\ are -540 and +580 km s$^{-1}$. {\bf(Bottom left):}The angle and the boundaries of the region summed (white dashed line), overlayed upon the 3-color {\em Spitzer} IRAC 3-color map and the \cplus\ contours (gray) and CO(1--0) contours (yellow).  {\bf(Bottom right):} The CO(1--0) PVD overlaid with a schematic representing the compact high velocity component (black), the splash ring (black) and regular rotation (white), as well as providing the positions of each region seen in Fig. \ref{fig:regions}, indicating the dominant molecular kinematic components represented in each.}
\label{fig:co_pv}
\end{figure*}

Figure \ref{fig:co_pv} shows the \cplus\ and CO(1--0) position-velocity diagram (PVD) of HCG~57a of a slice intersecting the disk (shown in the inset panel).  The \cplus\ emission and the CO(1--0) emission seem to have broadly similar distributions, but the \cplus\ PVD contains some contamination from HCG~57d  around 9000 km s$^{-1}$ because of the generous width chosen for the extraction box. The CO(1--0) map  suffers less contamination because of the smaller beam. In the CO(1--0), instead of a single velocity component, HCG~57a seems to have three.  

First, there is a regularly rotating component, seen as an S-shaped curve that begins in the lower left of the PVD in Figure \ref{fig:co_pv}, and terminates in the upper right, shown in white on the schematic.  This component spans the CO extent.

The second kinematic component in the PVD is seen at the top right, but has a smaller velocity extent, deemed the splash ring.  The splash ring seems to be a tight knot of molecular gas, seen in the velocity slices from 31--349 \kms\ (8758--9076 \kms) in the CO channel map shown in Figure \ref{fig:co_chans} on the right-hand side of the main velocity component within the HCG~57a disk.  The knot also appears to be correlated with extinction, which is visible in the underlying SDSS 3-color image.  

The third kinematic component spans a small spatial scale and a large velocity, ranging from -400 to +250 \kms\ (8277 to 8977 \kms), labeled the compact, high velocity component.  This broad component is offset from the peak of the CO(1--0) magnitude, which corresponds to the nucleus of the galaxy.  This kinematically broad and spatially compact kinematic component lacks the symmetry expected from a bar, which would be co-rotating with the disk.  It is therefore possible that this broad, compact kinematic component is either an inflow or an outflow.  While $\pm325$ \kms\ is large,  this component does not exceed the escape velocity.  \citet{lisenfeld+14} shows that the broad CO velocities in HCG~57a compared with $L_{K2.2\mu{\rm m}}$ of the galaxy make it an outlier compared with the other MOHEGs.  The high velocity feature is offset from the nucleus (measured using a 2MASS image; \citealt{2mass}) by $\approx4$\arcsec, meaning it is less likely to be AGN-driven outflow, and may be a tidal in origin\footnote{We note that HCG~57a shows only weak evidence of an active nucleus, having a marginally enhanced [N~{\sc ii}]/H$\alpha$ optical emission line ratio \citep{martinez+10}}.  

\citet{Torres+14} claim the detection of a faint H$\alpha$ structure connecting HCG~57d to the center of HCG~57a, and it is possible that the compact, high velocity component might be related.  The separation of these distinct CO components is not currently possible within the \cplus, due to the limitations of the spatial (9.4\arcsec) resolution of {\em Herschel}.  It is also possible that there is faint, diffuse CO emission associated with the \cplus\ from HCG~57d, but falls far below the detection threshold of CARMA and can be considered negligible.

Following the kinematic components of the PVD, HCG~57a was divided into three separate regions when constructing integrated CO(1--0) and \cplus\ line profiles, using the region boxes labeled on Figure \ref{fig:regions}.  Each region is dominated by a different molecular kinematic component.  R1, at the lower left corner of HCG~57a appears to only have the regular rotation component in the CO(1--0) PVD.  R2 encompasses the nuclear region of HCG~57a, which includes the compact, high velocity kinematic feature and regular rotation.  R3 is found at the top right, which includes the regular rotation component, and the splash ring component in the north-west of the PVD.

The boundaries of R2 were chosen from a careful exploration of the emission in the PACS data cube, to minimize obvious contamination of emission in \cplus\ from HCG~57d.  Figure \ref{fig:57a_specs} presents the \cplus\ and CO(1--0) spectra for each region, as well as an integrated spectrum of all regions.  The total luminosities for all regions were calculated by summing all channels inside of the shaded velocity ranges (blue for CO(1--0) and yellow for \cplus) on the plots in Fig. \ref{fig:57a_specs}.  Because of the complex structure of both the \cplus\ and CO(1--0), summing over all channels with emission was preferable to attempting to fit the spectra with Gaussian profiles.  Table \ref{tab:fluxes} presents the \cplus\ and CO(1--0) line luminosities of the regions of HCG~57a.  The \cplus/CO(1--0) ratio varies from region to region, ranging between 560--1380.  Unlike in the case of HCG~57d, these ratios do not fall outside the range for normal galaxies \citep{stacey+10}.


\section{Discussion}
\label{discussion}

The HCG~57 system was targeted by our team using {\it Herschel} to try to gain more understanding of systems with high warm-H$_2$/PAH ratios as derived from {\em Spitzer}, which are surmised to be shock-dominated systems. In particular, the disk of HCG~57a was shown to contain significant extended warm H$_2$ emission not explained by heating in PDRs associated with star formation sites \citep{cluver+13}.  Recently, \citet{appleton+13} have shown that in Stephan's Quintet, \cplus\ emission is also enhanced in  shocked regions, and that these regions are coincident with sites of significant kinetic energy dissipation, as evidenced by the existence of large quantities of broad-line  (600--700 km s$^{-1}$) CO-emitting molecular gas. The turbulent gas between the Taffy galaxies, UGC~12914/5, \citep{peterson+12} show similar properties (Peterson et al. in preparation).  Both Stephan's Quintet and the Taffy bridge are intergalactic regions, and serve as well-resolved examples of how shocks can heat molecular gas away from other sources of heating.  Although perhaps not as extreme as Stephan's Quintet, we find that HCG~57d has exceptional \cplus\ emission-line properties compared with normal galaxies. HCG~57a shows mildly enhanced \cplus\ properties and strong star formation suppression, especially in regions where the velocity field of the molecular gas is highly disturbed. We will discuss the possibility that these galaxies are experiencing boosted molecular and atomic gas heating (over and above that associated with star formation) associated with the recent collision.

\subsection{Testing the validity of PDR heating for the HCG~57 system}
\label{disc:pdr}

For galaxies in which O+B stars in star-forming regions are the dominant energy source heating the ISM, the \cplus\ arises primarily in PDRs associated with diffuse gas heated by photoelectric heating from polycyclic aromatic hydrocarbons (PAHs) and small dust grains \citep{watson72,draine78,tielens+85,bakes+94}.  [C~{\sc ii}] emission can also arise from gas collisionally heated by other sources, ranging from hot electrons in an ionized medium (H~{\sc ii} regions), shocks and/or turbulence \citep{appleton+13,lesaffre+13}, as well as X-ray Dissociation Regions (XDRs) or Cosmic Rays.  XDRs are unlikely in this case, as the X-ray flux reported for the HCG~57 system in \citet{cluver+13} was insufficient to power the warm H$_2$, and is likely not the dominant heating source of the \cplus\ either.

\begin{figure}[t!]
\includegraphics[width=0.49\textwidth]{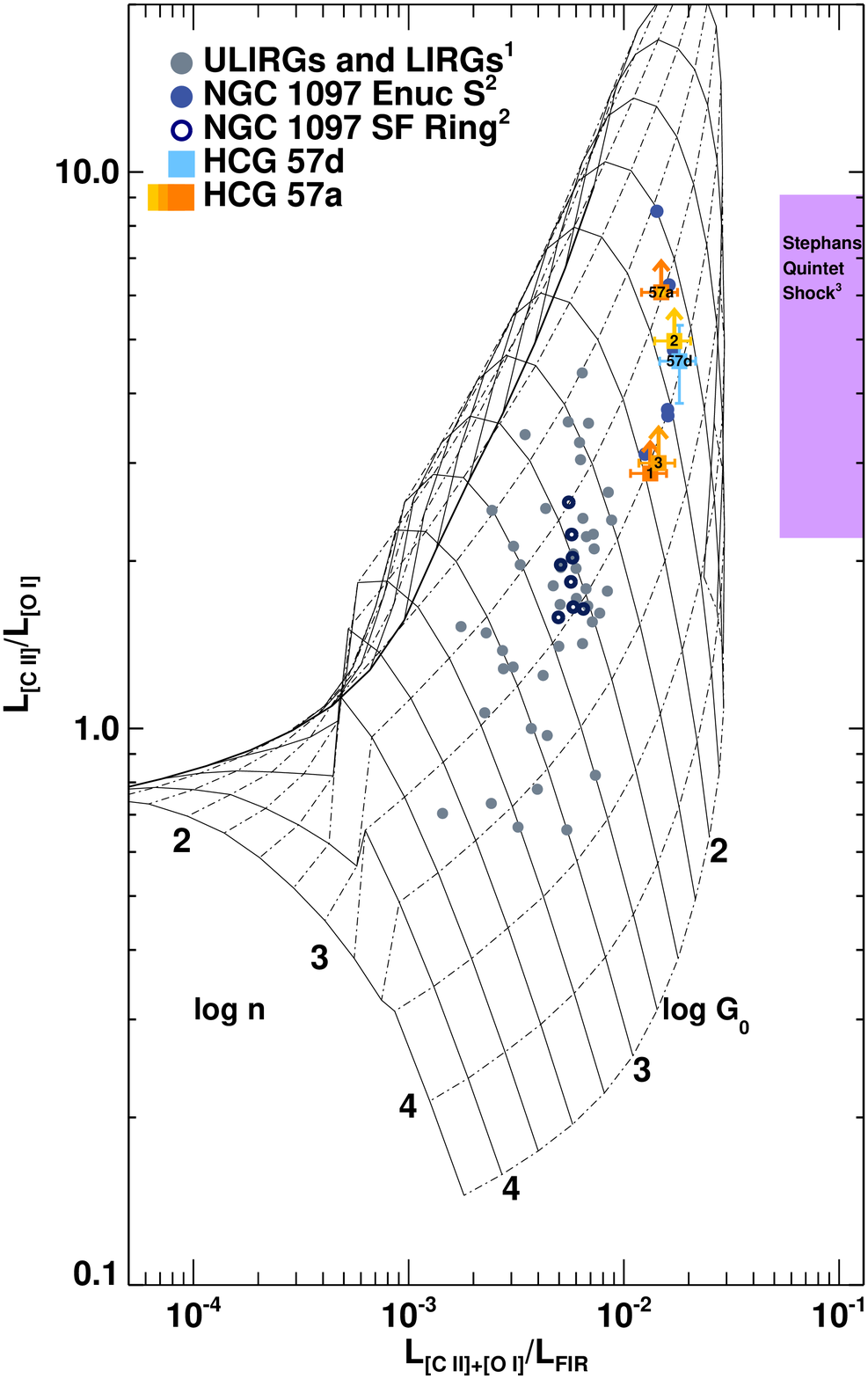}
\caption{The \cplus--to--\oi\ ratios plotted against the \cplus\ and \oi\ cooling compared to the far-infrared luminosity, overplotted on PDR models as well as data points from star-forming galaxies from 1:\citet{malhotra+01} and the resolved {\em Herschel} data of NGC~1097 from 2:\citet{beirao+12}.  Error bars include \cplus\ and \oi\ calibration uncertainties.  Extrapolating the expected density $n$, and incident radiation field $G_0$, indicates that if PDRs are the dominant \oi\ and \cplus\ excitation source for the HCG~57 system, then HCG~57a should be primarily low density and low $G_0$ with the emission in HCG~57d coming from a slightly higher density gas.  It is of note that the lowest density is expected in Region 2, with $n~\approx10^2$ particles per cm$^3$.  The range of \oi\ and \cplus\ emission properties of the Stephan's Quintet shock is shown as a purple box (2: \citealt{appleton+13}).}
\label{fig:oi_cii}
\end{figure}

Comparing the total luminosities in each line, we see that the \cplus/\oi\ ratio is 0.23 for HCG~57d, which is on the extreme lower end of the distribution for the integrated properties of normal galaxies \citep{malhotra+01}.   Figure \ref{fig:oi_cii} shows the \cplus/\oi\ ratio of HCG~57d as well as the lower limits of the ratio for Regions 1--3 in HCG~57a, compared with their far-IR line cooling, against PDR models of \citet{kaufman+99}, and the starburst ring and diffuse bar-regions (extra-nuclear region S) of NGC~1097 from \citet{beirao+12}.   We also show the data for normal galaxies from \citep{malhotra+01}. The points associated with HCG~57d and the three HCG~57a regions lie in the upper right-hand corner of the available PDR model parameter space. Region 2, which includes the most kinematically-disturbed gas is the most extreme of the HCG~57a individual regions plotted. These points would move further upwards had \oi\ been detected.  Based on these IR-line diagnostics alone, the PDR models would require low UV radiation fields (low $G_0$) and low densities ($n_{\rm H}<10^2$ cm$^{-3}$) and low average gas pressure ($P_{\rm gas}<10^4$~K~cm$^{-3}$). For reference, we also show the locus of observations from an  almost pure extragalactic shock from the Stephan's Quintet filament of \citet{appleton+13} as a purple box in  Figure \ref{fig:oi_cii}.  A mix of shocks combined with higher-density PDRs could therefore also explain the positions of these points on the figure. 

Several of the points from the southern diffuse bar region of NGC~1097 \citep{beirao+12} occupy similar regions to HCG~57a in Figure \ref{fig:oi_cii}.  These regions of NGC~1097 have long been associated with a strong radio continuum ``hook'' \citep{ondrechen+83} which is believed to correspond to diffuse gas and plasma compressed in a shock-wave associated with the stellar bar. More recent observations and models by \citet{beck+05} strongly support the idea that this region of NGC~1097 is dominated by highly compressed shocked gas associated with a narrow-dust lane, which is instrumental in feeding gas towards the inner part of the galaxy and starburst ring.  It is likely that the diffuse \cplus\ emitting gas in NGC~1097Õs southern region could also be boosted by shock-heating of molecular gas passing through the same compression region.  CO imaging studies of the bar region of NGC~1097 could shed light on how similar it is to the HCG~57 system.

\begin{figure}[t!]
\includegraphics[width=3.4in,clip,trim=1.1cm 0.2cm 0.8cm 1cm]{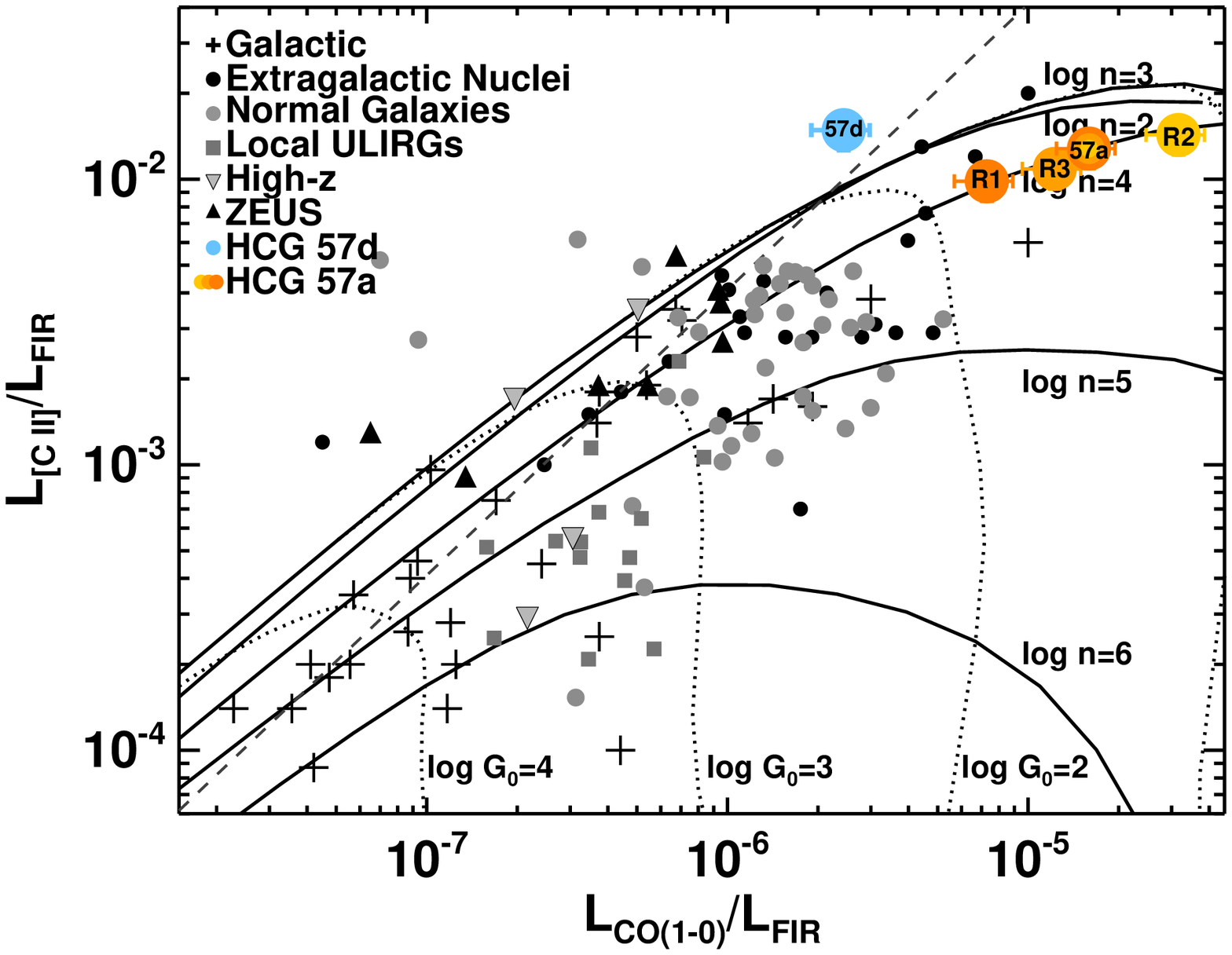}
\caption{The \cplus\ and CO(1--0) emission measurements of HCG~57 are plotted on the plot originally published in \citet{stacey+10}.  Overplotted are the PDR model values for as a function of n and G$_0$ from PDRT \citep{kaufman+99}.  Error bars represent the flux calibration uncertainties of {\em Herschel} and CARMA.  The \cplus/L$_{\rm FIR}$ ratio of HCG~57d is too high to be associated with PDRs.  In fact, the L$_{\cplus}$/L$_{\rm CO(1-0)}$ ratio exceeds theoretical maximum (of 4100; dashed black line) by a factor of $\approx1.5$.  This is an indication that a mechanism other than star formation are partially powering the \cplus\ emission in the galaxy.  If HCG~57a indeed has \cplus\ emission excited by shocks, then the region points due purely to PDRs would move down from the extreme diffuse gas and the low $G_0$ to parameters more physical for the amount of CO(1--0) detected by CARMA.}
\label{fig:stacey}
\end{figure}

Figure \ref{fig:stacey} plots the \cplus/$L_{\rm FIR}$ against the CO(1--0)/$L_{\rm FIR}$ of both galaxies, compared to normal and high redshift galaxies \citep{stacey+10}.  We again show, for reference, the PDR models of \citet{kaufman+99}, but this time including CO emission predictions. The points for both galaxies again lie in the upper far right of the plot, on average beyond the sources in the original \citet{stacey+10} work.  The plot shows that HCG~57d lies significantly above the locus of points for normal galaxies presented by \citet{stacey+10}, and above the ``maximum'' PDR limit.  HCG~57d is outside the extreme edge of a pure PDR model in Fig. \ref{fig:stacey}, despite being within the extreme range of PDR models in Figure \ref{fig:oi_cii}.  The host galaxy properties show emission consistent with that of a low-density and low $G_0$ PDR ($G_0<10^2$), but would likely require a boost from an additional collisional energy source, either shocks, ionized gas or a warm neutral (H~{\sc i}) medium to push it above the PDR limit.

For HCG~57a, the \oi, \cplus\ and CO(1--0) shown in Figures \ref{fig:oi_cii} and \ref{fig:stacey} are in direct disagreement, if we assume that photoelectric heating from PDRs is the dominant heat source in both the cold and warm molecular gas.  The lower limit to the \cplus/\oi\ ratio from Figure \ref{fig:oi_cii} indicates that HCG~57a requires a diffuse gas ($n\lesssim10^3$ cm$^{-3}$) and low incident $G_0$.  The CO(1--0) and \cplus\ also requires low $G_0$, though needs a density 1--2 orders of magnitude higher to explain the line ratios.  The disagreement between the models that fit the \cplus, \oi, and CO(1--0) suggests a more complex picture for the ISM in HCG~57a. In fact, HCG~57a may consist of two ISM components, a PDR-dominated component (the interface between young stars and cold gas)  and warmer diffuse component. We note that Region 2, which is the most extreme in Figures \ref{fig:oi_cii} and \ref{fig:stacey}, coincides with the position of the IRS slit from \citet{cluver+13}, and likely traces the same emitting area.  In this region, $L$(\cplus$)\sim L{\rm (H{_2,warm})}$, consistent with what would be predicted in diffuse shocked molecular gas \citep{appleton+13,lesaffre+13}.  \citet{cluver+13} indicates that the extended warm H$_2$ emission in HCG~57a requires an additional heating source, therefore it is likely that a portion of the \cplus\ is excited by a similar mechanism.  Shocks in the diffuse ISM would naturally explain the strong H$_2$ emission and possible boosting of [C~{\sc ii}].

Is it fair to compare the conditions in HCG~57a and 57d to that of the giant shocked filament in Stephan's Quintet? As \citet{appleton+13} speculated, the isolation of the shocked gas away from significant star formation allowed the effects of shocks to be explored with only minimal PDR contamination in the Quintet. As \citet{cluver+13} showed, the HCG galaxies with large warm H$_2$/PAH ratios also have specific star formation rates which are an order of magnitude lower than normal galaxies, and therefore we might expect that shocks and turbulence, if present, may have a larger proportional effect than in normal galaxies. We note that the HCG~57 system shares some similarities with the Stephan's Quintet  shock, namely 1) broad CO and \cplus\ lines, 2) enhanced \cplus/FIR emission, and 3) an overabundance of warm molecular hydrogen inconsistent with photoelectric heating by grains in a UV radiation environment. We therefore think that it is reasonable to consider galaxies like HCG~57a and 57d as candidates for shock-boosted warm interstellar gas.

\subsection{Can we rule out diffuse ionized gas as an extra source of [C~{\sc ii}] emission in HCG~57?}
To evaluate the importance of diffuse ionized gas from H~{\sc ii} regions, we searched for the presence of strong nebular lines, like the [N~{\sc ii}]205$\mu$m line \citep{beirao+12} via deep {\em Herschel} Spectral and Photometric Imagine Receiver (SPIRE) Fourier Transform Spectrograph (FTS; \citealt{spire}) observations. The [N~{\sc ii}]205$\mu$m line is an excellent tracer of ionized gas, because its ionization potential lies just above that of neutral hydrogen. The 3$\sigma$ upper limit for the [N~{\sc ii}]205$\mu$m line in HCG~57a is $<4.5\times10^{-18}$~W~m$^{-2}$ over the central beam (FWHM = 16.9\arcsec\ at $\nu$ = 1.410~THz) in a pointing centered on Region 2.  Scaling up the \cplus\ for Region 2 by a factor of 1.3 to correct for the larger SPIRE beam, we estimate the \cplus/[N~{\sc ii}]205$\mu$m ratio to be $>17$. This ratio is at least a factor of 3-5 $\times$ larger than that associated with H~{\sc ii} regions for a reasonable range of densities, and thus we conclude that in the center of HCG~57a, the ionized gas contribution to the \cplus\ emission is small. The SPIRE FTS footprint did not include HCG~57d, therefore we cannot rule out a contribution to the \cplus\ emission from H~{\sc ii} regions in that galaxy.  

Gas with low-metallicity can also lead to large values of \cplus/CO(1-0) ratio through an extended PDR, and this is often found in dwarf systems (e.g. \citealt{Israel+11}). In order to determine whether the \cplus/CO(1--0) ratio was due to such low-metallicity effects, we used the spectrum taken of HCG~57d by the Sloan Digital Sky Survey (SDSS).  Using Equation 11 from \citet{kewley+02}, log(O/H)+12 for HCG~57d = 8.648, which is close to solar abundance (HCG~57a does not have a SDSS spectrum available).  The fact that the metallicity is so close to solar means that it is unlikely that the \cplus/$L_{\rm FIR}$ ratio is discrepant due to low metallicity effects.  Although the optical fiber of HCG~57d only covers the central $3''$ of the galaxy, it is unlikely that such a small galaxy could experience a strong metallicity gradient.  Deep optical integral field spectroscopy of ionized gas in HCG~57d may further illuminate the reason for the boosted \cplus/$L_{\rm FIR}$, by providing spatially resolved ionized gas ratios, which are able to determine the dominant source of ionization \citep{kewley+06}.

\subsection{Suppressed star formation in HCG~57a?}
\label{supp_SF}
The molecular gas mass and surface densities of HCG~57a and HCG~57d were calculated based on the CO(1--0) map, convolved to a point spread function (PSF) of $9.4''$, and registered to the \cplus\ map, which was chosen to allow the regions to remain consistent throughout this paper.  The mass of H$_2$ for a given flux was determined assuming the standard $L_{\rm CO}$--to--M(H$_2$) conversion factor of \citep{bolatto+12}:
\[\frac{\rm M(H_2)}{\rm M_\odot} = 8.47\times10^3\left(\frac{D}{\rm Mpc}\right)^2 \left(\frac{\int S_\nu\Delta v}{\rm Jy~km~s^{-1}}\right)\]

\begin{figure}[t]
\includegraphics[width=3.4in]{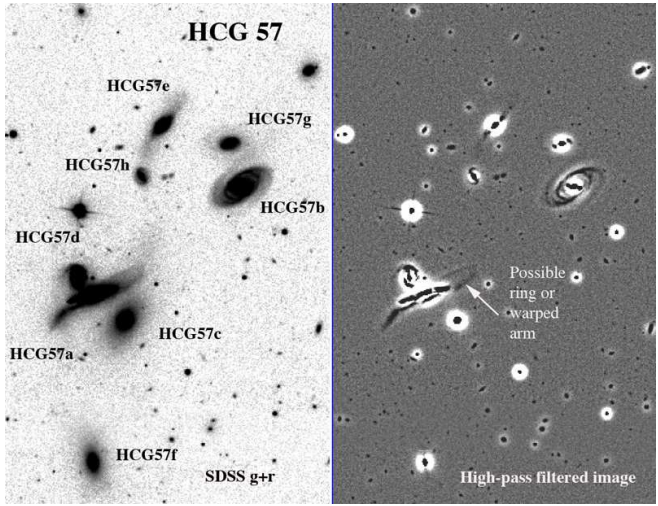}
\caption{{\bf(Left)}: The SDSS g and r images summed to bring out faint structure for the HCG~57 group. Our paper concentrates on the HCG~57a,d sub-system.  {\bf(Right)}: A high-pass filtered version of the same image after subtracting the image from a smoothed version of itself (smoothed with a gaussian with a $\sigma = 3 pixels$), which suppresses extended structure and emphasizes sharp features in the image, such as the spiral arms in HCG~57b, nuclear bar structures and the ring/ring-arcs in HCG~57a and d). }
\label{fig:unsharp}
\end{figure}

The total H$_2$ masses calculated for HCG~57a and 57d are listed in Table \ref{tab:co+sf}\footnote{We use the total H$_2$ mass from \citet{lisenfeld+14} for HCG~57d in the following analysis}, of $7.22\times10^9$~M$_\odot$ for HCG~57a and $2.28\times10^9$~M$_\odot$ for HCG~57d.  Using the stellar masses calculated in \citet{bitsakis+14} of $2.0\times10^{11}$~M$_\odot$ and $2.5\times10^{10}$~M$_\odot$ for HCG~57a and 57d, respectively, correspond to molecular gas-to-stellar mass fractions of 3.6\% and 9.1\%, respectively, which is in the normal range for late-type galaxies.  The lower limit to the molecular fraction\footnote{Molecular-to-neutral gas fraction} on the other hand is quite high.  The total group \hi\ mass from \citet{verdes-montenegro+01} is $5.1\times10^9$~M$_\odot$.  The HCG~57 group is comprised of two sub-groups (Fig. \ref{fig:unsharp}), and therefore the \hi\ mass listed is an upper limit to the HCG~57a, c, d complex within the group.  Including only M(H$_2$) derived from CO(1--0) in HCG~57a and 57d, we derive the lower limit to the molecular fraction of $f_{\rm mol} > 0.65$, and is likely more dominant if the \hi\ is distributed more uniformly across the group (which is likely).  A larger survey along the sequence of compact group \hi\ depletion \citep{verdes-montenegro+01} will shed light about whether there is a relationship between molecular gas fraction and \hi\ starvation (Alatalo et al. 2014, in preparation).

\begin{table}[b!]
\caption{HCG~57 CO(1--0) Fluxes, $\Sigma_{\rm H_2}$ and $\Sigma_{\rm SFR}$}
\centering
\begin{tabular*}{3.5in}{l c c c c}
\hline \hline
Name & S$_{\rm CO}$ & M(H$_2$)$^\dagger$ & log $\Sigma_{\rm mol}$ & log $\Sigma_{\rm SFR}$\\
& (Jy~km~s$^{-1}$) & $\times10^9$ (M$_\odot$) & (M$_\odot$~pc$^{-2}$) & (M$_\odot$~yr$^{-1}$~kpc$^{-2}$)\\
\hline
HCG~57d &
5.24$\ddagger\pm$1.1 &
0.78$\pm$0.16 &
0.94 & -2.15\\
HCG~57a &
48.6$\pm$2.4 &
7.22$\pm$0.35 &
1.54 & -2.53\\
Region~1 &
8.93$\pm$1.2 &
1.33$\pm$0.18 &
1.25 & -2.54\\
Region~2 &
26.9$\pm$1.1 &
3.99$\pm$0.16 &
1.73 & -2.44\\
Region~3 &
12.0$\pm$0.84 &
1.78$\pm$0.13 &
1.48 & -2.67\\

\hline \hline
\label{tab:co+sf}
\vspace{-3mm}
\end{tabular*}
\raggedright $^\dagger$Cold H$_2$ mass as measured by CO(1--0) and converted using $X_{\rm CO} = 2\times10^{20}$ cm$^{-2}$ (K km s$^{-1}$)$^{-1}$ \citep{bolatto+12} \\
$^\ddagger$ \citet{lisenfeld+14} recovered 10.2$\pm$2.1 Jy~km~s$^{-1}$ in HCG~57d, corresponding to a total molecular mass of $2.28\pm0.47\times10^9$~M$_\odot$.
\end{table}

The gas surface density $\Sigma_{\rm H_2}$ was then determined by dividing by the area in which CO(1--0) was detected, based on the areas shown in Fig. \ref{fig:regions}.  GALEX far-ultraviolet broadband and the {\em Spitzer} 24\micron\ maps were sky-subtracted using the IDL routine {\tt sky}\footnote{http://idlastro.gsfc.nasa.gov/ftp/pro/idlphot/sky.pro} from the NASA Goddard IDL Astronomy User's Library, then convolved to 9.4$''$ and registered to the \cplus\ map as well, to allow matching resolution and matching pixels for all maps used to determine the SF rate and gas surface density. The SF rate was then determined from the FUV+24\micron\ conversion from \citet{leroy+08}, normalized to a Salpeter Initial Mass Function \citep{salpeter55} (for comparison to the original \citet{ken98} sample).  The SFRs of HCG~57a and 57d estimated from this method agreed well with the SFRs derived in \citet{bitsakis+14}.  The SFR was averaged over the same pixels that had non-zero emission in the CO(1--0) moment0 map within each of the pertinent regions.  The CO fluxes, H$_2$ masses, and surface densities are listed in Table \ref{tab:co+sf}.

Figure \ref{fig:off_ks} shows the Kennicutt-Schmidt relation (K-S; \citealt{ken98}) of HCG~57d and the three regions of HCG~57a, compared to other objects including the Milky Way \citep{yusef-zadeh+09}, normal galaxies \citep{ken98,fisher+13}, high redshift objects \citep{genzel+10} and radio galaxies \citep{ogle+10}.  While HCG~57d appears to agree with the relation, HCG~57a appears to be suppressed by factor of 5--18.  In fact, the observed suppression seems to correlate with regions of the CO(1--0) PVD with multiple kinematic components along the line-of-sight (Fig. \ref{fig:co_pv}b).  R1, which sits nearest to the K-S relation, only contains one kinematic component, belonging to the rotating disk.  On the other hand, both R2 and R3 contain multiple kinematic components (the regular rotation, compact high velocity and splash ring components) along the line of sight, and are also the objects that appear the farthest from the K-S relation.

The fact that the suppressed regions of the galaxy correspond quite well to the regions with disturbed kinematics in the PVD points to the possibility that the interaction that caused the large velocity disturbances seen in the galaxy are also currently suppressing the SF in those regions.  The CO kinematics in HCG~57d and R1 are both regular, predicting that those regions would form stars normally, and indeed, it appears that they do.  In contrast, R2 and R3 have CO kinematics that do not indicate simple rotation alone.  Star formation requires fragmentation of the gaseous disk due to gravitational instabilities.  If another source of heating was available to the gas (such as turbulence), it could be changing the balance between kinetic energy and potential energy, and therefore reducing the efficiency of SF.  The large amount of rotationally excited H$_2$ seen in HCG~57a \citep{cluver+13}, as well as the high$\cplus$/FIR and $\cplus$/CO(1-0) ratios, points to the likelihood that an additional heating source operates in this galaxy: perhaps turbulence and shocks.

\citet{guillard+12a} argued in the context of SQ that a shock is able to inject turbulent energy and heat the molecular gas in the system, stabilizing it against collapse, and thus explain why we see such a low SFR in the most kinematically disturbed regions of HCG~57a.  If the \cplus\ enhancement and CO(1--0) kinematics are indeed due to shock dynamics, this is a natural explanation for the suppression of SF seen in HCG~57a.

On the other hand, the fact that the molecular gas that falls the farthest from the K-S relation tends to be the most kinematically disturbed is suggestive of another possibility for the suppression.  Namely, that assuming a particular L(CO)--M(H$_2$) relationship holds for most external systems, when the relation has been measured and constrained using virialized giant molecular clouds within the Milky Way \citep{solomon+87}.  If the underlying clouds in the kinematically disturbed portions of HCG~57a have a smoother distribution, as is seen in ultraluminous infrared galaxies (ULIRGS; \citealt{downes+98}), then X$_{\rm CO}$ could be a factor of $\approx2-3$ lower \citep{sandstrom+13}.  \citet{bitsakis+14} calculated the gas-to-dust ratios in the HCG~57 system, and found that the dust mass in HCG~57d to be $7.8\times10^6$~M$_\odot$ and HCG~57a to be $4.0\times10^7$~M$_\odot$.  If HCG~57a and 57d have a standard gas-to-dust ratio of $\approx 150$, consistent with that found in a large sample of HCGs (Alatalo et al., in preparation), as well as with the Taffy Galaxies \citep{zhu+07}, the gas mass derived using the \citet{bolatto+12} X$_{\rm CO}$ conversion is consistent within errors.


\subsection{Deciphering the interaction history of HCG~57a and 57d}
The enhanced \cplus\ emission, disturbed CO(1--0) kinematics and suppressed SF in the HCG~57 system point to an interesting recent interaction history between the HCG~57a and HCG~57d. We show in Figure \ref{fig:unsharp} an SDSS view the HCG~57 group as a whole (left = sum of SDSS g and r images and right = same image after application of a high-pass filter to bring out the narrow structures). It is clear from the images that HCG~57a and d show signs of interaction, and both system contain complete or partial ring-like structures.  Although we cannot rule out recent interaction with other group members, the most likely cause of the warped nature of HCG~57a, is its nearby companion HCG~57d.  

We hypothesize that HCG~57d, the smaller companion has recently splashed through HCG~57a, but did so off-center, creating a collisional ring \citep{appleton+sm06}  in HCG~57d, and inducing a strong ring disturbance in the edge-on galaxy HCG~57a. The offset between the center of the ring and the nucleus is 2 arcseconds ($0.3\times$ minor axis radius of the ring) as measured on the SDSS g image.  This is similar to, for example, the offset in the famous offset ring the ``Cartwheel'' and Arp 10 \citep{appleton+97, charmandaris+93}.  Models of off center collisions (e. g. \citealt{toomre78} -- Figure 5, \citealt{appleton+87}, and \citealt{gerber+94}) show that this kind of offset is created when the impact parameter between the two galaxies is roughly 1/4 of the radius of the disk.

This off-center collision hypothesis is also able to explain why the SDSS and {\em Spitzer} images from Figures \ref{fig:cii_co} and \ref{fig:cii_chans} show the nucleus of the smaller galaxy to be offset from the center of its star-forming ring, a common prediction for non-zero impact parameter ring galaxies \citep{appleton+sm06}. In this picture, both galaxies should evolve into galaxies with rings or ring-segments, depending how offset the collision is. The partial ring or warped arm seen in the high-pass filtered image of HCG~57a could be interpreted as such a collisionally driven ring-wave which is also warped by the passage of HCG~57d through the disk of HCG~57a.  

Although the above scenario is speculative, it might explain the peculiar kinematics of HCG~57a seen in the CARMA data,  especially the upper righthand corner of the PVD (Fig. \ref{fig:co_pv})  where the double line profiles could be interpreted as part of a radially expanding wave centered to the north-west of the galaxies center. The characteristic timescale of the high velocity PVD feature on the north-western side of HCG~57a is about 50~Myr (assuming an expansion velocity of 100 km s$^{-1}$ and a radius of 5 kpc),  which is shorter than the approximate rotational timescale of the galaxy of approximately 160 Myrs, implying that if this is the kinematic signature of an expanding wave, it has not fully disturbed the whole disk. Without a detailed model of the gas and stellar dynamics of a collision, it is hard to further quantify. Given that this is a quite dense association of galaxies, we cannot rule out other members of the group having also played a role. 

\begin{figure}[t!]
\includegraphics[width=3.4in,clip,trim=1.5cm 0.4cm 0.5cm 0.9cm]{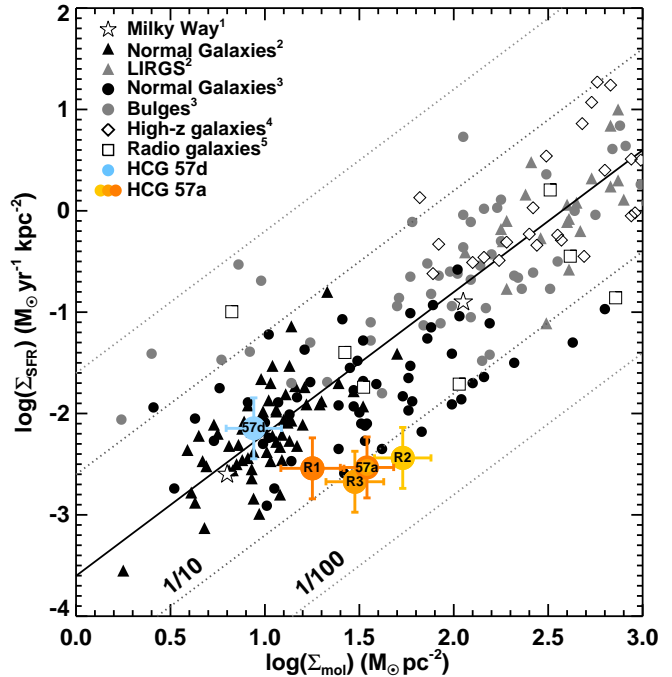}
\caption{Star formation and gas density within the HCG~57 system (derived in \S\ref{results}) are shown in comparison to their position on the Kennicutt-Schmidt relation \citep{ken98}, including HCG~57d, the integrated HCG~57a point as well as the regions.  The SFR was calculated using the 24\micron\ {\em Spitzer} data combined with the far-UV data from {\em GALEX}, using \citet{leroy+08}, normalized to a Salpeter Initial Mass Function (IMF).  Error bars represent the scatter in the $24\mu$m--to--SFR conversion \citep{calzetti+07}, and the calibration uncertainties and $L_{\rm CO}$--to--H$_2$ conversion.  HCG~57d and regions of HCG~57a are compared to SF-to-gas density for objects including the Milky Way (1: \citealt{yusef-zadeh+09}), normal galaxies and bulges (2: \citealt{ken98}, 3: \citealt{fisher+13}), high redshift objects (4: \citealt{genzel+10}) and radio galaxies (5: \citealt{ogle+10,alatalo+14}), all renormalized to a Salpeter IMF. The lines shown represent the Kennicutt-Schmidt relation, and 1/10 and 1/100 suppression and enhancement lines. }
\label{fig:off_ks}
\end{figure}

Even without a proper model of the proposed collision between HCG~57a and d, it is interesting to note that the most SF suppressed regions of HCG~57a (Regions 2 and 3 in Figure \ref{fig:off_ks}) are also the regions with the largest velocity disturbances (Fig. \ref{fig:co_pv}b). This is consistent with the results seen in Stephan's Quintet filament, where multiple broad CO velocity components, symptomatic of kinetic energy dissipation in the molecular gas, are found associated with low star formation rates \citep{guillard+12a,konstantopoulos+14}. It is possible therefore that strong kinematic disturbances, caused by galaxy collisions, can suppress star formation. 

We finally come to the question of why HCG~57a and d have such different \cplus/CO ratios. Our observations have shown that HCG~57d has received a boost in its \cplus\ emission relative to that expected from PDR models, and that HCG~57a could potentially also be enhanced, but to a lesser degree. One reason may have to do with the relative stellar masses of the two galaxies (of $2.0\times10^{11}$ M$_\odot$ for HCG~57a and $2.5\times10^{10}$ M$_\odot$ for HCG~57d; \citealt{bitsakis+14}). High values of \cplus\ to CO ratios are not uncommon in low-metallicity dwarf systems \citep{Poglitsch+95,stacey+91,Israel+96,Israel+11,Madden+97}, and in such systems the "dark-gas"  fraction (i.e.. molecular gas not traced by CO)  is expected to increase with lower metallicity \citep{wol10}. However, we believe that HCG~57d is closer to solar oxygen metallicity (see earlier), and thus the dark molecular fraction is likely to be modest.

Instead, we suggest that the collision between the two galaxies could create an enhancement in the diffuse component relative to the denser molecular gas by a different mechanism. HCG~57d, being the smaller of the two galaxies, would receive a significantly stronger perturbation than HCG~57a, and this would inject more kinetic energy into the ISM of that galaxy, as well as causing almost all the gas in the galaxy to be affected quickly because of its small size. Furthermore, if the galaxy has indeed crashed through the disk of HCG~57a, the diffuse gas in the galaxy would be significantly affected and heated. Although we do not have sufficient spatial resolution in the Herschel observations to determine whether the \cplus\ emission is much more extended than the denser molecular gas, one plausible explanation for the enhanced \cplus\ emission is that the collision excited a much larger volume of diffuse gas than the denser CO molecular gas, thus increasing the observed ratio.


\subsection{The HCG~57 system in context}
The gas-on-gas collision that HCG~57 has recently experienced are common occurrences in Hickson Compact Groups \citep{hickson97}, due to their high space densities and low galaxy velocity dispersions.  This environment predisposes galaxies to interact with one another, and is likely a conduit for rapid evolution, as suggested by the dearth of galaxies in the infrared (IR) green valley \citep{johnson+07,cluver+13}.  The MOHEG systems in particular seem to be in a special portion of their evolution, as these galaxies tend to be the ones found in the IR green valley \citep{cluver+13}, rapidly transitioning from blue and starforming to red and dead.  

The suppressed SF seen in HCG~57a might be a beacon pointing toward the physics that dictates how that transition takes place.  The scenario we put forward is that when two galaxies within the compact group have a direct collision, the gas is heated through shocks, suppressing star formation in the immediate term, due to the injection of the extra kinetic energy to the molecular gas, via turbulence \citep{guillard+09,guillard+12a,appleton+13}.  Shocks in general are shown to enhance SF \citep{jog+92,barnes+96,elmegreen+97}, but we hypothesize that this SF enhancement requires a sufficient amount of time for the cooling lines to efficiently shed the energy that was injected by turbulence, which effectively balanced gravity.  Once the turbulence has been efficiently cooled, the larger density of the post-shock gas determines the SF enhancement.  It is likely that the suppression that is seen in HCG~57a is transient, but speaks to the role that turbulence likely plays in inhibiting SF.  On longer scales, HCGs will also suffer from harassment \citep{icke85, mihos95, moore+96, bekki98} and ram pressure stripping (\citealt{cayatte+90,bohringer+94}; though this was argued against in \citealt{cluver+13}), which can remove the ISM, effectively shutting down future SF and transitioning the galaxy.

High redshift galaxies are known to contain more turbulent gas \citep{shapiro+09}, and therefore it is important to understand to what degree turbulence modifies SF, both by enhancing it and suppressing it.  Studying MOHEGs within HCGs is an ideal way to gauge the role of turbulence in environments already known to be extreme, and the HCG~57 system has shown itself to be an ideal test case.

\section{Summary}
\label{summary}
We presented the \oi, \cplus\ and CO(1--0) maps taken using the {\em Herschel} Space Observatory and the CARMA array of the Hickson compact group galaxies HCG~57a and HCG~57d, in which HCG~57a is a known MOHEG.

\begin{itemize}
\item The \cplus/L$_{\rm FIR}$ ratio in HCG~57d is too large to be explained with photoelectric heating in PDRs alone.  Two possible explanations for the boosted \cplus/$L_{\rm FIR}$ are copious amounts of ionized gas, or possibly the shock excitation of a larger volume of diffuse gas compared with the denser molecular gas resulting from HCG~57d having crashed through the disk of HCG~57a. Excitation of \cplus\ by neutral gas is unlikely because of the low observed \hi\ content of the HCG~57 group.

\item The \cplus/CO(1--0) ratio of HCG~57a is also more easily explained if one assumes that a shock has traversed the system, boosting the \cplus\ emission and resolving discrepancies in PDR models needed to explain the CO(1--0)/$L_{\rm FIR}$ and \cplus/$L_{\rm FIR}$ ratios.   SPIRE FTS upper limits to the detection of the [N~{\sc ii}]205$\mu$m line allow us to rule out copious ionized gas as the boosting agent in this case. The approximate equality in the cooling power of warm H$_2$ mid-IR emission and \cplus\ emission, detected in the center of HCG~57a, are consistent with models of  shocks or turbulence boosting the \cplus\ emission relative to PDR emission.

\item The CO(1--0) gas kinematics in HCG~57a shows complex structure, including regions with kinematic components beyond the simple galactic rotation curve.  The star formation is suppressed preferentially in those regions with complex kinematic structures, by factors of 30--40 (as compared with normal efficiencies in HCG~57d and R1, both of which have simple molecular gas kinematics). The dust-to-gas ratio in HCG~57a appears to be consistent with that seen in other HCGs using a ``standard'' $L_{\rm CO}$--M(H$_2$) conversion factor, and we hypothesize that the current suppression has been brought on by injected turbulence.

\item A collision between HCG~57a and HCG~57d 50 Myr ago, where HCG~57d directly intersected the disk of HCG~57a would possibly explain both the unusual \cplus\ properties, as well as the star formation suppression.  In HCG~57d, the collisionally-induced shock has already traversed the entire disk of the galaxy, and \cplus\ cooling has been enhanced, allowing for efficient star formation in the (now denser) postshock gas.  The shock, on the other hand, has not completely traversed HCG~57a, meaning that turbulent injection is still taking place in the disk.  This extra energy injection counteracts efficient fragmentation and gravitational collapse, therefore suppressing star formation. These processes may cause rapid termination of star formation, leading to rapid evolution of the galaxy through the mid-infrared green valley.
\end{itemize}

\acknowledgments K.A. would like to thank Drew Brisbin and Gordon Stacey for giving access to the data and code used to create the \cplus/CO comparison figure used in this paper, and the anonymous referee for insightful recommendations that have improved the paper.
K.A. and P.N.A. supported by funding through {\em Herschel}, a European Space Agency Cornerstone Mission with significant participation by NASA, through an award issued by JPL/Caltech. U.L. acknowledges  support by the research projects.   AYA2011-24728 from the Spanish Ministerio de Ciencia y Educaci\'on and the Junta de Andaluc\'\i a (Spain) grants FQM108.  T.B. and V.C. would like to acknowledge partial support from the EU FP7 Grant PIRSES-GA-2012-316788.  LVM work has been supported by grant AYA2011-30491-C02-01 co-financed by MICINN and FEDER funds, and the Junta de Andaluc\'ia (Spain) grants P08-FQM-4205 and TIC-114.
Support for CARMA construction was derived from the Gordon and Betty Moore Foundation, the Kenneth T. and Eileen L. Norris Foundation, the James S. McDonnell Foundation, the Associates of the California Institute of Technology, the University of Chicago, the states of California, Illinois, and Maryland, and the National Science Foundation. Ongoing CARMA development and operations are supported by the National Science Foundation under a cooperative agreement, and by the CARMA partner universities. {\em Herschel} is an ESA space observatory with science instruments provided by European-led Principal Investigator consortia and with important participation from NASA.  This work is based [in part] on observations made with the {\em Spitzer} Space Telescope, which is operated by the Jet Propulsion Laboratory, California Institute of Technology under a contract with NASA.
Funding for the SDSS and SDSS-II has been provided by the Alfred P. Sloan Foundation, the Participating Institutions, the National Science Foundation, the U.S. Department of Energy, the National Aeronautics and Space Administration, the Japanese Monbukagakusho, the Max Planck Society, and the Higher Education Funding Council for England. The SDSS Web Site is http://www.sdss.org/.

 

 \end{document}